# Differential Modulation in Massive MIMO With Low-Resolution ADCs

Don-Roberts Emenonye, *Student Member, IEEE*, Carl Dietrich, *Senior Member, IEEE*, and R. Michael Buehrer, *Fellow, IEEE*

***Abstract*—In this paper, we present a differential modulation and detection scheme for use in the uplink of a system with a large number of antennas at the base station, each equipped with low-resolution analog-to-digital converters (ADCs). We derive an expression for the maximum likelihood (ML) detector of a differentially encoded phase information symbol received by a base station operating in the low-resolution ADC regime. We also present an equal performing reduced complexity receiver for detecting the phase information. To increase the supported data rate, we also present a maximum likelihood expression to detect differential amplitude phase shift keying symbols with low-resolution ADCs. We note that the derived detectors are unable to detect the amplitude information. To overcome this limitation, we use the Bussgang Theorem and the Central Limit Theorem (CLT) to develop two detectors capable of detecting the amplitude information. We numerically show that while the first amplitude detector requires multiple quantization bits for acceptable performance, similar performance can be achieved using one-bit ADCs by grouping the receive antennas and employing variable quantization levels (VQL) across distinct antenna groups. We validate the performance of the proposed detectors through simulations and show a comparison with corresponding coherent detectors. Finally, we present a complexity analysis of the proposed low-resolution differential detectors.**

## I. Introduction

MASSIVE multi-input-multi-output (MIMO) is a core feature in fifth generation (5G) cellular communication [1]. Generally, a massive MIMO system has hundreds of antennas at the base station (BS), each equipped with independent radio-frequency (RF) chains. This configuration has the potential to provide large multiplexing and diversity gains [2]. Conditioned on the availability of channel state information at the transmitter (CSIT), this large dimensionality has the potential for the complete elimination of inter-user interference using matched filter beamforming for the downlink and uplink respectively [3].

Although the benefits of massive MIMO are abundant, two main challenges hinder its deployment. The first challenge concerns the prohibitively high circuit power incurred by the massive number of RF chains at the BS. A particular point of power inefficiency occurs at the ADCs. It has been shown that the total power consumed by the ADCs increases exponentially with an increase in their resolution [4]. This is a huge issue since much of the promise shown by early research in massive MIMO was contingent on infinite/high resolution analog-to-digital converters. A popular approach in solving the power consumption challenge is to employ low-resolution ADCs [5]–[8]. In [6], a one-bit distributed reception scheme is investigated for IoT devices. In that work, the channel is assumed to be perfectly known and maximum likelihood and zero-forcing-type receivers are derived. In [5], the problem of estimation with one bit is framed as a convex optimization problem. An iterative estimator for symbol-by-symbol estimation is provided. An algorithm for estimating the channel is also provided, where the number of required pilot symbols scales with the number of users. In [9], a MIMO-OFDM system with low-resolution ADCs is proposed, and a maximum a-posterior (MAP) algorithm for channel estimation and symbol detection is investigated. In [10], an amplify-and-forward massive MIMO system is developed. Using the Bussgang decomposition [11], channel estimation for the fully digital relay system is presented and exact expressions for the achievable rate are derived. It should be noted that in [10], [12], the number of pilots required for channel estimation scales with the number of antennas in the relaying system. In [13], a supervised learning approach is used to provide more reliable channel estimation and data detection in the low-resolution regime.

Although the application of low-resolution ADCs alleviates the issue of energy inefficiency, it adds an additional challenge. More specifically, in low-resolution systems, a large number of pilots is required for channel estimation. Hence, the channel is usually assumed to remain constant for a longer period of time [14]. In fast or frequency selective fading channels, this assumption presents a major limitation. This major limitation motivates the investigation of differential modulation in the low-resolution regime.

Even though differential modulation is well-developed in the general area of communications [15]–[18], there has been a resurgence of interest in the topic. This is partly because non-coherent/differential systems do not require instantaneous channel state information at either the transmitter or the receiver. In [19], [20], the authors show that any differential modulation scheme can be implemented using a look-up table. The look-up table is constructed by minimizing the non-coherent distance between two distinguishable codewords. Additionally, a two-symbol detector is also proposed and evaluated in [19], [20]. In [21], non-coherent modulation is employed and an autocorrelation-based decision-feedback

Don-Roberts Emenonye, Carl Dietrich and R. Michael Buehrer are with the Bradley Department of Electrical and Computer Engineering, Virginia Polytechnic Institute and State University, Blacksburg, VA, 24060 USA e-mail: (donroberts@vt.edu; rbuehrer@vt.edu).

differential detection technique is adopted. In [22], [23], a constellation design is proposed for non-coherent modulation and an energy detection scheme is presented for reception. In [24], a differential detection scheme is provided for 16-quadrature amplitude modulation (16-QAM) systems, in which adjacent symbols are used to detect the current symbol in the presence of an infinite number of antennas. The differential detection scheme for 16-APSK and 16-QAM is extended to the finite antenna regime scenario in [25]. In [26], a constellation design based on pulse amplitude modulation (PAM) is proposed, which minimizes the symbol-error rate when the channel statistics are known. In [26], the receiver is based on energy detection and its bit-error rate performance is derived for the high signal-to-noise ratio and the large number of antennas scenarios. This work is extended in [27] to a massive MIMO scheme where orthogonal codes are used to allow for multi-user transmissions. A new constellation is provided and the decision regions are optimized. Authors in [28] present a combination of differential modulation and 5G beam management procedures for use in the downlink of a MIMO-OFDM system. In [29], the impact of reconfigurable intelligent surfaces on non-coherent phase modulation is analyzed. Authors in [30] propose expectation propagation based detectors for non-coherent multi-user MIMO. *These prior works do not consider the impact of low-resolution ADCs. Thus, in this work we develop differential detectors for the uplink of a massive MIMO system where each receive antenna employs low-resolution ADCs.*

More specifically, in this work we present, for the first time, an analysis of the uplink of a massive MIMO system with differential modulation employed at the transmitter and low-resolution ADCs employed at the base station. We focus on linear differential shift keying systems presented in [19], [31]. The contributions of this work are as follows

- The first contribution of this work is to develop a closed-form expression for the maximum likelihood detector of a differential phase shift keying system with one-bit ADCs employed at each antenna at the base station. To aid this development, we assume that the effect of quantization is constant across the entire duration of transmissions. Subsequently, the quantized received signal at the current channel use is represented as a linear function of the quantized signal received at previous channel uses[1]. We show through numerical simulations that the detector derived from this expression has a comparatively good BER performance in the large number of antenna regime.
- Second, to avoid the numerical burden associated with the repeated computation of the standard normal CDF function and to reduce the detection complexity, we present a linear receiver[2].
- Third, we extend the system from differential phase shift keying modulation to differential amplitude phase shift keying modulation and present a receiver that detects amplitude and phase information. We note that although this detector can detect the phase information, it fails to accurately detect the amplitude information, leading to poor performance.
- Fourth, we increase the number of quantization bits and employ the Bussgang Theorem [11] to decompose the received quantized signal into two parts - a desired component and an uncorrelated quantization noise component. This expression is used to derive a low-resolution (2 bits) energy-based receiver capable of reliably detecting the amplitude information.
- Fifth, to derive a single bit receiver that detects amplitude information, we place the receive antennas into groups such that antennas in the same group employ the same quantization level and antennas in different groups employ different quantization levels. This setup is termed antenna grouping-based variable quantization level (VQL) setup. With this setup, the Bussgang Theorem and the CLT are used to develop an energy-based receiver for amplitude detection.
- Finally, we analyze the complexity of the proposed receivers and provide Monte-Carlo based performance comparisons with corresponding coherent receivers. The comparisons indicate that the proposed detectors can achieve similar BER performance with coherent detectors while attaining a better spectral efficiency.

*Notation:* Matrices are denoted by bold uppercase letters (i.e. $\boldsymbol{H}$), vectors are denoted by bold lowercase letters (i.e. $\boldsymbol{h}$), $\Re(\cdot)$ denotes the real part of the argument, $\Im(\cdot)$ denotes the imaginary part of the argument, $(\cdot)^T$ denotes the transpose operator, $(\cdot)^H$ denotes the hermitian transpose, $(\cdot)^*$ denotes the conjugate operation, $(\cdot)^\dagger$ denotes the matrix inverse, $\%$ represents the modulus operator, $|\cdot|$ specifies the absolute value of a complex argument. $|\cdot|$ also specifies the cardinality of a set, $\mathbf{1}_U$ denotes a $U \times 1$ all one vector, $\boldsymbol{I}_U$ denotes a $U \times U$ identity matrix, and $\|\cdot\|$ specifies a norm consistent with the two norm.

## II. System Model

We define the system model, the transmission and reception modes relevant to the detector designs. Across all transmission and reception modes, the base station is equipped with $U$ antennas each equipped with low-resolution ADCs. In this work, we consider both a single carrier system with a frequency-selective block fading channel model and an OFDM multi-carrier system experiencing both frequency-selective and fast fading channels. The frequency-selective and fast fading components of the channel are due to multipath and high doppler respectively. During the $n-$th symbol duration, the $u-$th base station antenna receives

$$\mathsf{y}_u[n] = \begin{cases} \sum_{l=0}^{L-1} \sum_{k=0}^{K-1} \mathsf{h}_{u,k}[n,l] \mathsf{x}_k[n-l] + \mathsf{z}_u[n], & \text{if OFDM,} \\ \sum_{k=0}^{K-1} \mathsf{h}_{u,k}[n] \mathsf{x}_k[n] + \mathsf{z}_u[n], & \text{if SC,} \end{cases} \tag{1}$$

where $\mathsf{x}_k[n]$ is the transmit signal from antenna $k$ with power $\mathbb{E}[|\mathsf{x}_k[n]|^2] = 1/K$, $\mathsf{z}_u[n] \sim \mathcal{CN}(0, \sigma_z^2)$ is a random variable modeling the base station impairments local to receive antenna

[1]In this article, both single-carrier and OFDM systems are considered. In the single-carrier system, a channel use is a single symbol interval and in OFDM, a channel use represents one subcarrier.

[2]A detector derived based on the normal CDF often leads to computationally expensive numerical algorithms that require high arithmetic precision [9].

$u$, and $\mathsf{h}_{u,k}[n,l]$ is the channel between the receive antenna $u$ and the transmit antenna $k$. The $l-$th multipath component at the $n-$th symbol duration for the multi-carrier system can be defined as

$$\mathsf{h}_{u,k}[n,l] = p[l] g_{u,k}[l] e^{j\left(\frac{2\pi f_l n}{N}\right)}, \tag{2}$$

where $p[l]$ is the power of the $l-$th tap with $\sum_{l=0}^{L-1} p[l] = 1$, $g_{u,k} \sim \mathcal{CN}(0,1)$, is the complex channel gain from the $k-$th transmitter to the $u-$th receive antenna of the $l-$th path, $f_l$ is the doppler spread of the $l-$th tap with $f_l = \frac{v_l}{\lambda}\cos\theta_l$, and $\theta_l$ is angle of arrival of the corresponding tap assumed to be uniformly distributed between $[-\pi,\pi]$. For single carrier, the frequency-selective channel at the $n-$th symbol duration can be modeled as a combination of $l$ parallel frequency-flat subchannels expressed as

$$h_{u,k}[v] = \sum_{l=0}^{L-1} p[l] g_{u,k}[l] e^{-j2\pi l v/N}. \tag{3}$$

For the transmission of $N$ symbols, the signal transmitted during the $n-$th symbol duration at the $k-$th transmit antenna can be defined as

$$\mathsf{x}_k[n] = \begin{cases} \frac{1}{\sqrt{N}} \sum_{v=0}^{N-1} s_k[v] e^{j2\pi n v/N}, & \text{if OFDM}, \\ s_k[n], & \text{if single-carrier}, \end{cases} \tag{4}$$

where $\mathsf{s}_k[n]$ is the data symbol. The data symbol is a zero-mean random variable with unit variance, i.e, $\mathbb{E}[\mathsf{s}_k[n]] = 0$ and $\mathbb{E}[|\mathsf{s}_k[n]|^2] = 1$. A cyclic prefix of length $N_{cp}$ is attached at the beginning of each OFDM symbol[3]

$$\mathsf{x}_k[n] = \mathsf{x}_k[N+n], \quad -N_{cp} < n < 0. \tag{5}$$

After taking an FFT and the removal of the cyclic prefix, the received signal has the following linear frequency domain representation[4]

$$y_u[v] = \sum_{k=0}^{K-1} h_{u,k}[v] x_k[v] + z_u[v], \tag{6}$$

where

$$x_k[v] = \frac{1}{\sqrt{N}} \sum_{n=0}^{N-1} \mathsf{x}_k[n] e^{-j2\pi n v/N}, \tag{7}$$

$$y_u[v] = \frac{1}{\sqrt{N}} \sum_{n=0}^{N-1} \mathsf{y}_u[n] e^{-j2\pi n v/N}, \tag{8}$$

$$h_{u,k}[v] = \frac{1}{\sqrt{N}} \sum_{l=0}^{L-1} \mathsf{h}_{u,l}[n] e^{-j2\pi l v/N}, \tag{9}$$

and $z_u[v] \sim \mathcal{CN}(0,\sigma_z^2)$ is the Fourier transform of the thermal noise local to the receive antenna $u$.

[3]For the single-carrier representation $n$ is equivalent to $v$

[4]While this ignores the fast-fading and inter-carrier interference resulting from high Doppler, these factors are not ignored during the Monte-Carlo simulation of the proposed low-resolution detectors.

### A. Quantization

After reception at each base station antenna, the real and imaginary part of the signal are independently quantized by identical low-resolution ADCs. To elaborate we define the quantization function as $q = \mathcal{Q}(s)$, where $\mathcal{Q}: \mathbb{C} \to \mathcal{E}_c$, and $\mathcal{E}_c$ is the set of complex quantization alphabets for the complex-valued inputs, $s$. The quantizer output can be decomposed as $q = q^R + jq^I$, where $q^R$ and $q^I$ represents the real-valued output corresponding to the real and imaginary part of the input $s$ respectively. The set of real-valued quantizer outputs is identical for both the real and the imaginary component. More specifically, $q^R, q^I \in \mathcal{E} = \{e_0, e_1, \cdots, e_{E-1}\}$, where $E = 2^{q_b}$ is the number of possible output values with $q_b$ being the number of ADC bits. To correctly represent the complex input sample, the quantizer needs $2q_b$ bits. With this, the set of complex quantization alphabets can be written as the Cartesian products, $\mathcal{E}_c = \mathcal{E} \times \mathcal{E}$. Clearly, the pair of labels $(q^R, q^I)$ is a quantized representation of the input complex variable $(s^R, s^I)$. These labels can be generated by comparing both $s^R$ and $s^I$ with $N_\zeta$ quantization bin boundaries

$$-\infty = \zeta_0 < \zeta_1 < \cdots < \zeta_{E-1} < \zeta_{N_\zeta} = +\infty.$$

Specifically, if the quantizer output is

$$q = \mathcal{Q}(s^R + s^I) = (\zeta_{l^R}, \zeta_{l^I}) \tag{10}$$

then $l^R \in \{0,1,\cdots,N_\zeta - 1\}$ and it satisfies $\zeta_{l^R} \leq s^R < \zeta_{l^R+1}$. Similarly, $l^I \in \{0,1,\cdots,N_\zeta - 1\}$ and satisfies $\zeta_{l^I} \leq s^I < \zeta_{l^I+1}$. A special case of the low-resolution quantization is the use of independent and identical one-bit ADCs to quantize the real and imaginary component of the input $s$. In this case, the quantization function can be represented by the $\mathrm{sgn}$ function and the quantization operation can be specified as

$$q = \mathrm{sgn}(\Re(s)) + j\,\mathrm{sgn}(\Im(s))), \tag{11}$$

where $\mathrm{sgn}$ is defined as

$$\mathrm{sgn}(\Re(s)) = \begin{cases} +1, & \text{if } \Re(s) \geq 0 \\ -1, & \Re(s) < 0 \end{cases}. \tag{12}$$

Hence, we have

$$q \in \{1+j, 1-j, -1+j, -1-j\}. \tag{13}$$

Taking advantage of the Bussgang theorem [11], the non-linear quantization effect of low-resolution quantizers can be decomposed into

$$q = \eta x + \epsilon, \tag{14}$$

where the $\eta$ is the quantization scaling factor and $\epsilon$ is the quantization noise component which is uncorrelated to $\eta$. Note that $\eta$ is chosen to minimize the error variance $\sigma_\epsilon^2 = \mathbb{E}[|\epsilon|^2]$ which is minimized by

$$\eta = \frac{\mathbb{E}[x^* q]}{\mathbb{E}[|x|^2]}. \tag{15}$$

## III. LOW-RESOLUTION DETECTION FOR DPSK

The differential encoding scheme considered for low-resolution detection was first presented in [31]. To adapt this scheme for low-resolution detection, we consider the transmission of $N$ symbols divided into $N/N_s$ sub-modular symbols of length $N_s$. Each of these sub-modular blocks are used to create $N_d \times N_d$ data matrices. Without loss of generality, we only consider square orthogonal data matrices described as

$$\boldsymbol{S}[v^{'}] = \frac{1}{\sqrt{N_s}} \sum_{n_s=0}^{N_s-1} \boldsymbol{A}_{n_s} s[v^{'}N_s + n_s] + \boldsymbol{B}_{n_s} s^{*}[v^{'}N_s + n_s], \tag{16}$$

where $s[v^{'}N_s + n_s]$ represents the $n_s$ element of the $v^{'}$ sub-modular block $\boldsymbol{s}[v'] \in \mathbb{C}^{N_s}$, $\boldsymbol{A}_{n_s}$ and $\boldsymbol{B}_{n_s}$ are $N_d \times N_d$ orthogonal matrices. These matrices are also orthogonal to each other. Note that $s[v^{'}N_s + n_s] \in \mathcal{S}$ where $\mathcal{S}$ is a phase shift keying constellation of size $M$. The transmission rate can be defined as $R = N_s/N_d$. It is well known that a data rate of 1 is only achievable for two transmit antennas.

To begin differential modulation, we define the differentially encoded matrix as

$$\boldsymbol{C}[v^{'}] = \begin{bmatrix} c_1[v^{'}N_d] & \dots & c_1[v^{'}N_d + N_d - 1] \\ \vdots & \vdots & \vdots \\ c_K[v^{'}N_d] & \dots & c_K[v^{'}N_d + N_d - 1] \end{bmatrix}, \tag{17}$$

where $\boldsymbol{C}[v^{'}] \in \mathbb{C}^{K \times N_d}$, $c[v^{'}N_d + n_d]$ represents the $n_d$ element of the $v^{'}$ sub-modular data block $\boldsymbol{c}[v'] \in \mathbb{C}^{N_d}$. The transmission sequence is initialized as $\boldsymbol{C}[-1] = \boldsymbol{I}_{K \times N_d}$ and subsequent differentially encoded matrices can be derived as

$$\boldsymbol{C}[v^{'}] = \boldsymbol{C}[v^{'} - 1]\boldsymbol{S}[v^{'}] \quad v^{'} = 0, 1, \cdots, N/N_d - 1. \tag{18}$$

The $N/N_d$ differentially encoded matrices are concatenated to form an $K \times N$ matrix, this operation is described as

$$\boldsymbol{C} = [\boldsymbol{C}[0], \boldsymbol{C}[1], \cdots, \boldsymbol{C}[N/N_d - 1]]. \tag{19}$$

The actual transmission matrix can be defined as

$$\boldsymbol{X} = \begin{cases} \boldsymbol{C}\boldsymbol{G}^H, & \text{if OFDM,} \\ \boldsymbol{C}, & \text{if single-carrier,} \end{cases} \tag{20}$$

where $G$ is the $N \times N$ discrete Fourier transform matrix. Note that $x_k[n]$ in (4) is the $(k, n)$ element of $\boldsymbol{X}$. The received signal specified by (4) and (6) for single carrier and OFDM respectively can be divided into $N/N_d$ groups. The $n^{'}-$th group of received signal can be specified as

$$\mathsf{y}_u[n^{'}] = \mathsf{X}^T[n^{'}]\mathsf{h}_u[n^{'}] + \mathsf{z}_u[n^{'}] \tag{21}$$

where

$$\begin{aligned} \mathsf{y}_u[n^{'}] &= [\mathsf{y}_u[n^{'}N_d], \mathsf{y}_u[n^{'}N_d + 1], \cdots, \mathsf{y}_u[n^{'}N_d + N_d - 1]]^T, \\ \mathsf{h}_u[n^{'}] &= [\mathsf{h}_{u,1}[n^{'}], \mathsf{h}_{u,2}[n^{'}], \cdots, \mathsf{h}_{u,K}[n^{'}]]^T, \\ \mathsf{z}_u[v^{'}] &= [\mathsf{z}_u[n^{'}N_d], \mathsf{z}_u[n^{'}N_d + 1], \cdots, \mathsf{z}_u[n^{'}N_d + N_d - 1]]^T, \end{aligned} \tag{22}$$

and

$$\mathsf{X}[n^{'}] = \begin{bmatrix} \mathsf{x}_1[n^{'}N_d] & \dots & \mathsf{x}_1[n^{'}N_d + N_d - 1] \\ \vdots & \vdots & \vdots \\ \mathsf{x}_K[n^{'}N_d] & \dots & \mathsf{x}_K[n^{'}N_d + N_d - 1] \end{bmatrix}. \tag{23}$$

Note that the SC representation can obtained by replacing $v$ by $n$. After an FFT operation for the OFDM system, the received signal can be analyzed in the frequency domain as

$$\boldsymbol{y}_u[v^{'}] = \boldsymbol{X}^T[v^{'}]\boldsymbol{h}_u[v^{'}] + \boldsymbol{z}_u[v^{'}]. \tag{24}$$

**Assumption 1.** *The channel is slowly varying and is approximately constant across two adjacent transmission matrices* $v^{'}$ *and* $v^{'} - 1$.

With Assumption 1, the received signal can be written as

$$\boldsymbol{y}_u[v^{'}] = \boldsymbol{S}^T[v^{'}]\boldsymbol{X}^T[v^{'} - 1]\boldsymbol{h}_u[v^{'} - 1] + \boldsymbol{z}_u[v^{'}], \tag{25}$$

and

$$\boldsymbol{y}_u[v^{'}] = \boldsymbol{S}^T[v^{'}]\boldsymbol{y}_u[v^{'} - 1] + \boldsymbol{z}^{'}_u[v^{'}], \tag{26}$$

where $\boldsymbol{z}^{'}_u[v^{'}] = \boldsymbol{z}_u[v^{'}] - \boldsymbol{S}^T[v^{'}]\boldsymbol{z}_u[v^{'} - 1]$. Note that $\boldsymbol{z}^{'}_u[v^{'}] \sim \mathcal{CN}(0, 2\sigma_z^2)$. The quantized version of the received signal at the $u-$th antenna during the $v^{'}$ transmission block can be written as

$$\boldsymbol{q}_u[v^{'}] = [q_u[v^{'}N_d], q_u[v^{'}N_d + 1], \cdots, q_u[v^{'}N_d + N_d - 1]]. \tag{27}$$

Through the Bussgang relation (14), the received signal can be written as

$$\frac{1}{\eta_{v'}}\boldsymbol{q}_u[v^{'}] - \frac{\epsilon_{v'}}{\eta_{v'}} = \boldsymbol{S}^T[v^{'}]\left[\frac{1}{\eta_{v'-1}}\boldsymbol{q}_u[v^{'} - 1] - \frac{\epsilon_{v'-1}}{\eta_{v'-1}}\right] + \boldsymbol{z}^{'}_u[v^{'}], \tag{28}$$

where $\rho_{v'}$ and $\epsilon_{v'}$ are the quantization gain and quantization error during reception of the $v^{'}$ transmission matrix.

**Assumption 2.** *The quantizer gain and quantizer noise remains constant during the entire transmission. The quantization noise is Gaussian.*

With Assumption 2, the quantized received signal can be written as

$$\begin{aligned} \boldsymbol{q}_u[v^{'}] &= \boldsymbol{S}^T[v^{'}]\boldsymbol{q}_u[v^{'} - 1] - \boldsymbol{S}^T\epsilon + \eta\boldsymbol{z}^{'}_u[v^{'}] + \epsilon, \\ \boldsymbol{q}_u[v^{'}] &= \boldsymbol{S}^T[v^{'}]\boldsymbol{q}_u[v^{'} - 1] + \boldsymbol{w}_u[v^{'}], \end{aligned} \tag{29}$$

where $\boldsymbol{w}_u[v^{'}] = \eta\boldsymbol{z}^{'}_u[v^{'}] - \boldsymbol{S}^T\epsilon + \epsilon$ is the combined effect of the thermal noise and the quantization noise, i.e, $\boldsymbol{w}_u[v^{'}] \sim \mathcal{CN}(0, \sigma_w^2 = 2\eta^2\sigma_z^2 + 2\sigma_\epsilon^2)$. $\boldsymbol{q}_u[v^{'}]$ can be written as

$$\begin{aligned} \boldsymbol{q}_u[v^{'}] = \frac{1}{\sqrt{N_s}} \sum_{n_s=0}^{N_s-1} & \boldsymbol{A}^T_{n_s}\boldsymbol{q}_u[v^{'} - 1]s[v^{'}N_s + n_s] \\ & + \boldsymbol{B}^T_{n_s}\boldsymbol{q}_u[v^{'} - 1]s^{*}[v^{'}N_s + n_s] + \boldsymbol{w}_u[v^{'}]. \end{aligned} \tag{30}$$

Defining the following:

$$\begin{aligned} \tilde{\boldsymbol{a}}_{u,n_s} &\triangleq \boldsymbol{A}^T_{n_s}\boldsymbol{q}_u[v^{'} - 1] \in \mathbb{C}^{N_d \times 1}, \\ \tilde{\boldsymbol{b}}_{u,n_s} &\triangleq \boldsymbol{B}^T_{n_s}\boldsymbol{q}_u[v^{'} - 1] \in \mathbb{C}^{N_d \times 1}, \\ \tilde{\boldsymbol{A}}_u &\triangleq \left[\tilde{\boldsymbol{a}}_{u,0}, \tilde{\boldsymbol{a}}_{u,1}, \cdots, \tilde{\boldsymbol{a}}_{u,N_s-1}\right] \in \mathbb{C}^{N_d \times Ns} \\ \tilde{\boldsymbol{B}}_u &\triangleq \left[\tilde{\boldsymbol{b}}_{u,0}, \tilde{\boldsymbol{b}}_{u,1}, \cdots, \tilde{\boldsymbol{b}}_{u,N_s-1}\right] \in \mathbb{C}^{N_d \times Ns}, \end{aligned} \tag{31}$$

and with these, (30) can be written as

$$\boldsymbol{q}_u[v^{'}] = {N_s}^{-0.5}[\tilde{\boldsymbol{A}}_u \boldsymbol{s}[v^{'}] + \tilde{\boldsymbol{B}}_u \boldsymbol{s}^*[v^{'}]] + \boldsymbol{w}_u[v^{'}]. \tag{32}$$

### A. One-Bit Detector For Single Carrier Systems

Taking the $l-$th element of $\boldsymbol{q}_u[v^{'}]$, we have

$$q_u[v^{'} N_d + l] = {N_s}^{-0.5}[\tilde{\mathbf{a}}^T_{u,l} \boldsymbol{s}[v^{'}] + \tilde{\mathbf{b}}^T_{u,l} \boldsymbol{s}^*[v^{'}]] + w_u[v^{'} N_d + l], \tag{33}$$

where $\tilde{\mathbf{a}}_{u,l} \in \mathbb{C}^{N_s}$ and $\tilde{\mathbf{b}}_{u,l} \in \mathbb{C}^{N_s}$ are the $l-$th row of $\tilde{\boldsymbol{A}}_u$ and $\tilde{\boldsymbol{B}}_u$ respectively. We can stack the vectors as

$$\begin{aligned} \boldsymbol{f}_u[l] &\triangleq [\tilde{\mathbf{a}}^T_{u,l} \;\; \tilde{\mathbf{b}}^T_{u,l}]^T \in \mathbb{C}^{2N_s}, \\ \tilde{\mathbf{s}}[v^{'}] &\triangleq [\boldsymbol{s}^T[v^{'}] \;\; \boldsymbol{s}^H[v^{'}]]^T \in \mathbb{C}^{2N_s}, \end{aligned} \tag{34}$$

and now (33) can be written as

$$q_u[v^{'} N_d + l] = {N_s}^{-0.5} \boldsymbol{f}^T_u[l] \tilde{\mathbf{s}}[v^{'}] + w_u[v^{'} N_d + l]. \tag{35}$$

To facilitate the derivation of the likelihood function, we transform the system model from the complex domain to the real domain. First, the quantized received signals are converted from the complex to the real domain

$$q_{R,u}[v^{'} N_d + l] = \begin{bmatrix} q_{R,u,1}[v^{'} N_d + l] \\ q_{R,u,2}[v^{'} N_d + l] \end{bmatrix} = \begin{bmatrix} \Re(q_u[v^{'} N_d + l]) \\ \Im(q_u[v^{'} N_d + l]) \end{bmatrix}, \tag{36}$$

next, $\boldsymbol{f}_u[l]$ at the $u-$th base station antenna can be written as

$$\boldsymbol{F}_{R,u}[l] = \begin{bmatrix} \Re(\boldsymbol{f}_u[l]) & \Im(\boldsymbol{f}_u[l]) \\ -\Im(\boldsymbol{f}_u[l]) & \Re(\boldsymbol{f}_u[l]) \end{bmatrix}^T = \begin{bmatrix} \boldsymbol{f}^T_{R,u,1}[l] \\ \boldsymbol{f}^T_{R,u,2}[l] \end{bmatrix} \in \mathbb{R}^{2 \times 4N_s}. \tag{37}$$

The base station refines[5] $\boldsymbol{f}_u[l]$ as

$$\widetilde{\boldsymbol{F}}_{R,u}[l] = \begin{bmatrix} \widetilde{\boldsymbol{f}}^T_{R,u,1}[l] \\ \widetilde{\boldsymbol{f}}^T_{R,u,2}[l] \end{bmatrix}, \tag{38}$$

where $\widetilde{\boldsymbol{f}}^T_{R,u,i}[l]$ is defined as

$$\widetilde{\boldsymbol{f}}^T_{R,u,i}[l] = q_u[v^{'} N_d + l] \boldsymbol{f}^T_{R,u,i}[l]. \tag{39}$$

The noise and the transmit signal can be written as

$$\begin{aligned} \boldsymbol{w}_{R,u}[l] &= \begin{bmatrix} \Re(w_u[l]) \\ \Im(w_u[l]) \end{bmatrix} = \begin{bmatrix} w_{R,u,1}[l] \\ w_{R,u,2}[l] \end{bmatrix} \in \mathbb{R}^{2 \times 1}, \\ \tilde{\boldsymbol{s}}_R[v^{'}] &= \begin{bmatrix} \Re(\tilde{\boldsymbol{s}}[v^{'}]) \\ \Im(\tilde{\boldsymbol{s}}[v^{'}]) \end{bmatrix} \in \mathbb{R}^{4N_s \times 1}. \end{aligned}$$

Now, we define two sets of indices, $\mathcal{P}$ and $\mathcal{N}$ based on the values of the vector $\boldsymbol{q}_u[v^{'}]$

$$\begin{aligned} \mathcal{P} &= \{(i, u, l) : q_u[v^{'} N_d + l] \geq 0\}, \\ \mathcal{N} &= \{(i, u, l) : q_u[v^{'} N_d + l] < 0\}. \end{aligned}$$

[5]The refinement operation is a sign change that allows for a compact representation of the one-bit likelihood function. The refinement operation only flips the sign of the previously received signal, if the current received signal is $-1$.

Finally, with these definitions, we can write the likelihood function as

$$\begin{aligned} & L(\tilde{\boldsymbol{s}}_R[v^{'}]) \\ =& Pr\left( \sqrt{\rho} \boldsymbol{f}^T_{R,u,i}[l] \tilde{\boldsymbol{s}}_R[v^{'}] + w_{R,u,i}[l] \geq 0 \middle| \forall (i,u,l) \in \mathcal{P} \right) \\ & Pr\left( \sqrt{\rho} \boldsymbol{f}^T_{R,u,i}[l] \tilde{\boldsymbol{s}}_R[v^{'}] + w_{R,u,i}[l] < 0 \middle| \forall (i,u,l) \in \mathcal{N} \right) \\ \overset{(a)}{=}& Pr\left( \sqrt{\rho} \tilde{\boldsymbol{f}}^T_{R,u,i}[l] \tilde{\boldsymbol{s}}_R[v^{'}] \geq -w_{R,u,i}[l] \middle| \forall (i,u,l) \in \mathcal{P} \right), \\ & Pr\left( \sqrt{\rho} \tilde{\boldsymbol{f}}^T_{R,u,i}[l] \tilde{\boldsymbol{s}}_R[v^{'}] \geq w_{R,u,i}[l] \middle| \forall (i,u,l) \in \mathcal{N} \right) \\ \overset{(b)}{=}& \prod_{i=1}^{2} \prod_{u=1}^{U} \prod_{l=0}^{N_s-1} \Phi\left( \sqrt{\rho} \tilde{\boldsymbol{f}}^T_{R,u,i}[l] \tilde{\boldsymbol{s}}_R[v^{'}] \right) \end{aligned} \tag{40}$$

where $\Phi(t) = \int_{-\infty}^{t} \frac{1}{\sqrt{2\pi}} \exp \frac{-\tau^2}{2} d\tau$, $\rho$ is the signal-to-noise ratio defined as $\rho = \frac{1}{\sigma^2_w}$, (a) is based on the sign refinement (38), and (b) is due to the fact that $w_{R,u,i}[l]$ is independent for all $i$, $l$ and $u$. This likelihood is similar to the likelihood in the coherent literature [6]. Hence, the ML detection rule is

$$\hat{\boldsymbol{s}}_R[v^{'}] = \underset{\tilde{\boldsymbol{s}}^{'}_R[n^{'}] \in \mathcal{S}_{R,N_s}}{\arg\max} \prod_{i=1}^{2} \prod_{u=1}^{U} \prod_{l=0}^{N_s-1} \Phi\left( \sqrt{\rho} \tilde{\boldsymbol{f}}^T_{R,u,i}[l] \tilde{\boldsymbol{s}}^{'}_R[v^{'}] \right). \tag{41}$$

Note that $\mathcal{S}_{R,N_s}$ is a constellation of phase shift keying symbols in the real domain with size $2M^{N_s}$.

**Remark 1.** *The likelihood functions of the differentially encoded information symbols in this paper are always derived conditioned on both the differentially encoded information symbol and the quantized signal received during the previous channel use or during the previous block of channel uses. If other conditions are needed for their derivation, they are stated.*

### B. General Low-resolution Detectors for Both OFDM and Single Carrier Systems

The complexity of the detector presented in (41) is exponential in regards to both $N_s$ and $M$. Also, the derivation assumes that the quantized signal is defined as in (13). This definition is only available for one-bit quantizers. Moreover, this definition doesn't describe the quantized received signal after the FFT operation. To design a general low-resolution detector, we restate (32):

$$\boldsymbol{q}_u[v^{'}] = {N_s}^{-0.5}[\tilde{\boldsymbol{A}}_u \boldsymbol{s}[v^{'}] + \tilde{\boldsymbol{B}}_u \boldsymbol{s}^*[v^{'}]] + \boldsymbol{w}_u[v^{'}]. \tag{42}$$

With this equation, and the following definitions $\tilde{\boldsymbol{q}}[v^{'}] \triangleq [\boldsymbol{q}^T_u[v^{'}] \;\; \boldsymbol{q}^H_u[v^{'}]]^T$ and $\tilde{\boldsymbol{w}}_u[v^{'}] \triangleq [\boldsymbol{w}^T_u[v^{'}] \;\; \boldsymbol{w}^H_u[v^{'}]]^T$, we have

$$\tilde{\boldsymbol{q}}_u[v^{'}] = {N_s}^{-0.5} \boldsymbol{\Gamma}_u \tilde{\boldsymbol{s}}[v^{'}] + \tilde{\boldsymbol{w}}[v^{'}], \tag{43}$$

where

$$\boldsymbol{\Gamma}_u = \begin{bmatrix} \tilde{\boldsymbol{A}}_u & \tilde{\boldsymbol{B}}_u \\ \tilde{\boldsymbol{B}}^*_u & \tilde{\boldsymbol{A}}^*_u \end{bmatrix} \in \mathbb{C}^{2N_d \times 2N_s}. \tag{44}$$

The composite noise is still a complex Gaussian variable, $\tilde{\boldsymbol{w}}[v^{'}] \sim \mathcal{CN}(0, \sigma_w^2)$, hence a maximum likelihood detector can be obtained as $\left\| \boldsymbol{q}_u[v^{'}] - {N_s}^{-0.5} \boldsymbol{\Gamma}_u \tilde{\boldsymbol{s}}[v^{'}] \right\|$. Note that this detector still has an exponential complexity. Now, we use the orthogonal nature of the encoding matrices to simplify the design of the detector. First, we note that

$$\begin{aligned} &\boldsymbol{\Gamma}_u^H \boldsymbol{\Gamma}_u \\ &= \begin{bmatrix} \tilde{\boldsymbol{A}}_u^H \tilde{\boldsymbol{A}}_u + \tilde{\boldsymbol{B}}_u^T \tilde{\boldsymbol{B}}_u^* & \tilde{\boldsymbol{A}}_u^H \tilde{\boldsymbol{B}}_u + \tilde{\boldsymbol{B}}_u^T \tilde{\boldsymbol{A}}_u^* \\ \tilde{\boldsymbol{B}}_u^H \tilde{\boldsymbol{A}}_u + \tilde{\boldsymbol{A}}_u^T \tilde{\boldsymbol{B}}_u^* & \tilde{\boldsymbol{A}}_u^T \tilde{\boldsymbol{A}}_u^* + \tilde{\boldsymbol{B}}_u^H \tilde{\boldsymbol{B}}_u \end{bmatrix} \in \mathbb{C}^{2N_s \times 2N_s}. \end{aligned} \tag{45}$$

Due to the orthogonal structure of the matrices, the $(n_s, n_s^{'})-$th element of the first diagonal element of $\boldsymbol{\Gamma}_u^H \boldsymbol{\Gamma}_u$ is

$$\begin{aligned} &\tilde{\boldsymbol{a}}_{u,n_s}^H \tilde{\boldsymbol{a}}_{u,n_s^{'}} + \tilde{\boldsymbol{b}}_{u,n_s}^T \tilde{\boldsymbol{b}}_{u,n_s^{'}}^* = \\ &\boldsymbol{q}_u^H[v^{'}-1] \boldsymbol{A}_{n_s}^* \boldsymbol{A}_{n_s^{'}}^T \boldsymbol{q}_u[v^{'}-1] + \boldsymbol{q}_u^T[v^{'}-1] \boldsymbol{B}_{n_s} \boldsymbol{B}_{n_s^{'}}^H \boldsymbol{q}_u^*[v^{'}-1] \\ &= \left\| \boldsymbol{q}_u[v^{'}-1] \right\|^2 \delta(n_s - n_s^{'}), \end{aligned} \tag{46}$$

and analysing the second diagonal component also reduces to $\left\| \boldsymbol{q}_u[v^{'}-1] \right\|^2 \delta(n_s - n_s^{'})$, hence

$$\tilde{\boldsymbol{A}}_u^H \tilde{\boldsymbol{A}}_u + \tilde{\boldsymbol{B}}_u^T \tilde{\boldsymbol{B}}_u^* = \tilde{\boldsymbol{A}}_u^T \tilde{\boldsymbol{A}}_u^* + \tilde{\boldsymbol{B}}_u^H \tilde{\boldsymbol{B}}_u = \left\| \boldsymbol{q}_u[v^{'}-1] \right\|^2 \boldsymbol{I}_{N_s}. \tag{47}$$

Because the two matrices $\boldsymbol{A}_{n_s}$ and $\boldsymbol{B}_{n_s}$ are orthogonal to each other for all $n_s$, the $(n_s, n_s^{'})$-th element of the first off-diagonal component of $\boldsymbol{\Gamma}_u^H \boldsymbol{\Gamma}_u$ gives

$$\begin{aligned} &\tilde{\boldsymbol{a}}_{u,n_s}^H \tilde{\boldsymbol{b}}_{u,n_s^{'}} + \tilde{\boldsymbol{b}}_{u,n_s}^T \tilde{\boldsymbol{a}}_{u,n_s^{'}}^* = \\ &\boldsymbol{q}_u^H[v^{'}-1] \boldsymbol{A}_{n_s}^* \boldsymbol{B}_{n_s^{'}}^T \boldsymbol{q}_u[v^{'}-1] + \boldsymbol{q}_u^T[v^{'}-1] \boldsymbol{B}_{n_s} \boldsymbol{A}_{n_s^{'}}^H \boldsymbol{q}_u^*[v^{'}-1] \\ &= 0, \end{aligned} \tag{48}$$

similarly, analyzing the $(n_s, n_s^{'})$-th element of the second off-diagonal component gives

$$\begin{aligned} &\tilde{\boldsymbol{b}}_{u,n_s}^H \tilde{\boldsymbol{a}}_{u,n_s^{'}} + \tilde{\boldsymbol{a}}_{u,n_s}^T \tilde{\boldsymbol{b}}_{u,n_s^{'}}^* = \\ &\boldsymbol{q}_u^H[v^{'}-1] \boldsymbol{B}_{n_s}^* \boldsymbol{A}_{n_s^{'}}^T \boldsymbol{q}_u[v^{'}-1] + \boldsymbol{q}_u^T[v^{'}-1] \boldsymbol{A}_{n_s} \boldsymbol{B}_{n_s^{'}}^H \boldsymbol{q}_u^*[v^{'}-1] \\ &= 0. \end{aligned} \tag{49}$$

Hence, the composite matrix can be written as

$$\boldsymbol{\Gamma}_u^H \boldsymbol{\Gamma}_u = \left\| \boldsymbol{q}_u[v^{'}-1] \right\|^2 \boldsymbol{I}_{2N_s \times 2N_s}. \tag{50}$$

Next, we transform (43) and decouple and the $N_s$ transmitted symbols:

$$\begin{aligned} &\tilde{\boldsymbol{r}}_u[v^{'}] \\ &= \boldsymbol{\Gamma}_u^H \tilde{\boldsymbol{q}}_u[v^{'}] = {N_s}^{-0.5} \left\| \boldsymbol{q}_u[v^{'}-1] \right\|^2 \boldsymbol{I}_{2N_s \times 2N_s} \tilde{\boldsymbol{s}}[n^{'}] + \tilde{\boldsymbol{\omega}}[v^{'}], \end{aligned} \tag{51}$$

where $\tilde{\boldsymbol{\omega}}[v^{'}] = \boldsymbol{\Gamma}_u^H \tilde{\boldsymbol{w}}[v^{'}]$. Note that $\tilde{\boldsymbol{r}}_u[v^{'}] = [\boldsymbol{r}_u^T[v^{'}] \boldsymbol{r}_u^H[v^{'}]]^T$ and $\tilde{\boldsymbol{\omega}}[v^{'}] = [\boldsymbol{\omega}[v^{'}]^T, \boldsymbol{\omega}[v^{'}]^H]^T$. Therefore, the second $N_s$ elements are conjugates of the first $N_s$ elements, hence the following equation can be used for detection

$$\boldsymbol{r}_u[v^{'}] = {N_s}^{-0.5} \left\| \boldsymbol{q}_u[v^{'}-1] \right\|^2 \boldsymbol{s}[v^{'}] + \boldsymbol{\omega}[v^{'}], \tag{52}$$

the detection consist of $N_s$ distinct detection rules

$$\begin{aligned} &\hat{s}[v^{'} N_s + l] = \\ &\underset{s^{'}[v^{'} N_s + l] \in \mathcal{S}}{\arg\min} \left\| r_u[v^{'} N_s + l] - {N_s}^{-0.5} \left\| \boldsymbol{q}_u[v^{'}-1] \right\|^2 s^{'}[v^{'} N_s + l] \right\|, \\ &l = 0, 1, \cdots, N_s - 1. \end{aligned} \tag{53}$$

## IV. Quantized Detectors for Differential Amplitude Phase Shift Keying

To differentially encode symbols using both phase and amplitude, we employ the classical differential amplitude phase shift keying [24] and we focus on a single-carrier setup. In this encoding scheme, two concentric circles are used, with each circle restricted to distinct amplitude levels $\{\psi_0, \psi_1\}$. The ring ratio between the two circles is defined as $a = \frac{\psi_1}{\psi_0}$, $\psi_0 = \sqrt{\frac{2}{a^2+1}}$, and the unit power constraint is maintained by $\psi_1^2 + \psi_0^2 = 2$. Similar to phase modulation, each point on the circles represents distinct phase shift keying symbols which are drawn from $\mathcal{S}$ In this section, a single antenna transmitter is considered, during the $v-$th symbol duration, the transmitter decides which of the available $2M$ constellation points to transmit. This decision is based on a block of $N_b$ bits $\boldsymbol{b}[v] = [b_1[v], b_2[v], b_3[v], \cdots, b_{N_b}[v]]$, and the transmitted symbol during the previous interval, $x[v-1]$. More specifically, the first bit, $b_1[v]$ determines the amplitude of the transmitted symbols, and the other $N_b - 1$ bits determine the phase. Suppose that the function $\Upsilon$ converts the rightmost $N_b - 1$ bits to an alphabet from the $M-$PSK constellation, $s[v] = \Upsilon([b_2[v], b_3[v], \cdots, b_{N_b}[v]])$, the differential encoding operation can be specified as

$$\begin{aligned} a[0] &= \psi[0], c[0] = a[0], \\ c[v] &= c[v-1] s[v], \\ x[v] &= a[v] c[v], \\ a[v] &= \begin{cases} 1, & \text{if } b_1[v] = 0, \\ \frac{\psi_1}{\psi_0}, & \text{if } b_1[v] = 1 \text{ and } \tilde{x}[v-1] = \psi_0 \ , \\ \frac{\psi_0}{\psi_1}, & \text{if } b_1[v] = 1 \text{ and } \tilde{x}[v-1] = \psi_1 \ , \end{cases} \end{aligned} \tag{54}$$

where $a[v] \in \mathcal{A}\{1, \frac{\psi_0}{\psi_1}, \frac{\psi_1}{\psi_0}\}$, $\tilde{x}[v] = |x[v]|$, and $\tilde{x}[v] \in \mathcal{X}\{\psi_0, \psi_1\}$. The transmitted symbol switches between the two concentric circles when the first bit is 1, otherwise, a the amplitude of the transmitted symbol remains constant across adjacent symbol intervals. Clearly, $c[v]$ and $s[v]$ are similar to the block matrices used for differential phase modulation with $N_d = N_s = 1$.

Considering the single carrier system, during the $v-$th symbol duration, the base station antenna $u$ receives:

$$y_u[v] = a[v] h_u[v] c[v] + z_u[v], \tag{55}$$

with Assumption 1 this equation becomes

$$\begin{aligned} y_u[v] &= \left( \frac{a[v]}{a[v-1]} \right) y_u[v-1] s[v] + z_u^{'}[v], \\ y_u[v] &= a^{'}[v] y_u[v-1] s[v] + z_u^{'}[v], \end{aligned} \tag{56}$$

where $z_u^{'}[v] = z_u[v] - a^{'}[v] s[v] z_u[v-1]$. Note that $z_u^{'}[v] \sim \mathcal{CN}(0, \varrho_z)$, where $\varrho_z = 2\sigma_z^2$ if $b_1[v] = 0$. Likewise, if $b_1[v] =$ 1, then $\varrho_z = \sigma_z^2(1+\frac{\psi_0^2}{\psi_1^2})$, or $\varrho_z = \sigma_z^2(1+\frac{\psi_1^2}{\psi_0^2})$. The signal-to-noise ratio translates to $\rho = \frac{1}{\varrho_z}$, and we define the amplitude ratio as $a^{'}[v] = \frac{a[v]}{a[v-1]}$. The received signal across all antennas can be written in the vectorized form

$$\boldsymbol{y}[v] = a^{'}[v]\boldsymbol{y}[v-1]s[v] + \boldsymbol{z}^{'}[v]. \tag{57}$$

After quantization, and using (14), the received signal can be written as

$$\frac{1}{\eta_v}q_u[v] - \frac{\epsilon_v}{\eta_v} = a^{'}[v]s[v]\left[\frac{1}{\eta_{v-1}}q_u[v-1] - \frac{\epsilon_{v-1}}{\eta_{v-1}}\right] + z_u^{'}[v]. \tag{58}$$

Similarly to the derivation for the phase detector, we use Assumption 2 to write the previous expression as

$$\begin{aligned} q_u[v] &= a^{'}[v]s[v]q_u[v-1] - a^{'}[v]s[v]\epsilon + \epsilon + \eta z_u^{'}[v], \\ q_u[v^{'}] &= a^{'}[v]s[v]q_u[v-1] + w_u[v], \end{aligned} \tag{59}$$

where $w_u[v] = \eta z_u^{'}[v] - a^{'}[v]s[v]\epsilon + \epsilon$ is the combined effect of the thermal noise and the quantization noise. Note that $z_u^{'}[v] \sim \mathcal{CN}(0, \varrho_{z,\epsilon})$, where $\varrho_{z,\epsilon} = \eta^2\varrho_z + 2\sigma_\epsilon^2$ if $b_1[v] = 0$. Likewise, if $b_1[v] = 1$, then $\varrho_{z,\epsilon} = \eta^2\varrho_z + \sigma_\epsilon^2(1+\frac{\psi_0^2}{\psi_1^2})$, or $\varrho_{z,\epsilon} = \eta^2\varrho_z + \sigma_\epsilon^2(1+\frac{\psi_1^2}{\psi_0^2})$. The signal-to-noise ratio translates to $\rho = \frac{1}{\varrho_{z,\epsilon}^2}$.

To facilitate the derivation of the likelihood function, we transform the system model from the complex domain to the real domain. First, the quantized and unquantized received signals are converted from the complex to the real domain

$$\boldsymbol{q}_{R,u}[v] = \begin{bmatrix} q_{R,u,1}[v] \\ q_{R,u,2}[v] \end{bmatrix} = \begin{bmatrix} \Re(q_u[v]) \\ \Im(q_u[v]) \end{bmatrix}, \tag{60}$$

$$\boldsymbol{y}_{R,u}[v] = \begin{bmatrix} y_{R,u,1}[v] \\ y_{R,u,2}[v] \end{bmatrix} = \begin{bmatrix} \Re(y_u[v]) \\ \Im(y_u[v]) \end{bmatrix}, \tag{61}$$

the channel between the transmitter and the $u-$th receiver, during the $v-$th symbol duration can be written as

$$\boldsymbol{H}_{R,u}[v] = \begin{bmatrix} \Re(h_u[v]) & \Im(h_u[v]) \\ -\Im(h_u[v]) & \Re(h_u[v]) \end{bmatrix}^T = \begin{bmatrix} \boldsymbol{h}_{R,u,1}^T[v] \\ \boldsymbol{h}_{R,u,2}^T[v] \end{bmatrix} \in \mathbb{R}^{2\times 2}. \tag{62}$$

Next, the quantized signal at the previous symbol interval received at the $u-$th base station antenna is converted from the complex to the real domain

$$\begin{aligned} \boldsymbol{F}_{R,u}[v] &= \begin{bmatrix} \Re(q_u[v-1]) & \Im(q_u[v-1]) \\ -\Im(q_u[v-1]) & \Re(q_u[v-1]) \end{bmatrix}^T \\ &= \begin{bmatrix} \boldsymbol{f}_{R,u,1}^T[v] \\ \boldsymbol{f}_{R,u,2}^T[v] \end{bmatrix} \in \mathbb{R}^{2\times 2}, \end{aligned} \tag{63}$$

and the base station refines[6] $\boldsymbol{f}_u[v]$ as

$$\widetilde{\boldsymbol{F}}_{R,u}[v] = \begin{bmatrix} \widetilde{\boldsymbol{f}}_{R,u,1}^T[v] \\ \widetilde{\boldsymbol{f}}_{R,u,2}^T[v] \end{bmatrix}, \tag{64}$$

[6]The refinement operation is a sign change that allows for a compact representation of the one-bit likelihood function. The refinement operation only flips the sign of the previously received signal, if the current received signal is $-1$.

where $\widetilde{\boldsymbol{f}}_{R,u,i}^T[v]$ is defined as

$$\widetilde{\boldsymbol{f}}_{R,u,i}^T[v] = q_u[v]\boldsymbol{f}_{R,u,i}^T[v]. \tag{65}$$

The noise and the transmit signal can be written as

$$\begin{aligned} \boldsymbol{w}_{R,u}[v] &= \begin{bmatrix} \Re(w_u[v]) \\ \Im(w_u[v]) \end{bmatrix} = \begin{bmatrix} w_{R,u,1}[v] \\ w_{R,u,2}[v] \end{bmatrix} \in \mathbb{R}^{2\times 1}, \\ \boldsymbol{s}_R[v] &= \begin{bmatrix} \Re(s[v]) \\ \Im(s[v]) \end{bmatrix} \in \mathbb{R}^{2\times 1}. \end{aligned}$$

Finally, the quantized received signal can be written as

$$\boldsymbol{q}_{R,u}[v] = a^{'}[v]\widetilde{\boldsymbol{F}}_{R,u}[v]\boldsymbol{s}_R[v] + \boldsymbol{w}_{R,u}[v], \tag{66}$$

and assuming that the SNR is known, the likelihood can be written as

$$L(a^{'}[v]\boldsymbol{s}_R[v]|\rho) = \prod_{i=1}^{2}\prod_{u=1}^{U}\Phi\left(a^{'}[v]\sqrt{\rho}\widetilde{\boldsymbol{f}}_{R,u,i}^T[v]\boldsymbol{s}_R[v]\right). \tag{67}$$

Note that the previous steps are contingent on the implicit assumption that $\rho$ is known at the receiver. This assumption is plausible for a purely phase shift keying differential scheme. However for a DAPSK scheme with an amplitude that is dependent on the transmitted bits, this assumption does not hold. As discussed previously, there are three variations of $\rho$, hence the likelihood can be written as

$$\begin{aligned} L(a^{'}[v]\boldsymbol{s}_R[v]|\rho) = \max\{&L(a^{'}[v]\boldsymbol{s}_R[v]|\rho_1), L(a^{'}[v]\boldsymbol{s}_R[v]|\rho_2), \\ &L(a^{'}[v]\boldsymbol{s}_R[v]|\rho_3)\}, \end{aligned} \tag{68}$$

where $\rho_1 = 1/(\eta^2\varrho_z + 2\sigma_\epsilon^2)$, $\rho_2 = 1/(\eta^2\varrho_z + \sigma_\epsilon^2(1+\frac{\psi_0^2}{\psi_1^2}))$, $\rho_3 = 1/(\eta^2\varrho_z + \sigma_\epsilon^2(1+\frac{\psi_1^2}{\psi_0^2}))$, and the maximum likelihood detector can be written as

$$\hat{a}^{'}[v]\hat{\boldsymbol{s}}_R[v] = \underset{(a^{'})^{'}[v]\boldsymbol{s}_R^{'}[v]\in\{\mathcal{A}\times S_R\}}{\arg\max} L((a^{'})^{'}[v]\boldsymbol{s}_R^{'}[v]|\rho). \tag{69}$$

### *A. Inverse-Decoding Receiver*

In line with the literature of low-resolution ADCs where the exact quantization model is avoided [9], we develop an inverse-decoding approach that aims to maximize an upper bound on the likelihood function. To facilitate the detector design, we collect the signals received during the $v-$th symbol duration

$$\boldsymbol{q}_R[v] = [\boldsymbol{q}_{R,1}^T[v],\ \ \boldsymbol{q}_{R,2}^T[v], \cdots, \boldsymbol{q}_{R,U}^T[v]]^T. \tag{70}$$

Next, we define the collection of the quantized signals received at the previous time step as $\widetilde{\boldsymbol{F}}_R[v] \in \mathbb{C}^{2U\times 2}$

$$\widetilde{\boldsymbol{F}}_R[v] = [\widetilde{\boldsymbol{F}}_{R,1}^T[v],\ \ \widetilde{\boldsymbol{F}}_{R,2}^T[v], \cdots, \widetilde{\boldsymbol{F}}_{R,U}^T[v]]^T, \tag{71}$$

and stack the arguments of the likelihood function

$$\begin{aligned} &\boldsymbol{\beta}(\boldsymbol{s}_R[v]) = [\beta_1(\boldsymbol{s}_R[v]),\ \ \beta_2(\boldsymbol{s}_R[v]), \cdots, \beta_{2U}(\boldsymbol{s}_R[v])]^T \\ &\beta_l(\boldsymbol{s}_R[v]) = \widetilde{\boldsymbol{f}}_{R,u,i}^T[v]\boldsymbol{s}_R[v], \end{aligned} \tag{72}$$

where $l = 2(u-1)+i$ for $u = 1, \cdots, U$ and $i = 1, 2$, the likelihood function can be written as

$$\begin{aligned} L(\boldsymbol{s}_R[v]) &= \prod_{i=1}^{2}\prod_{u=1}^{U} \Phi(\sqrt{\rho}\widetilde{\boldsymbol{f}}_{R,u,i}^T[v]\boldsymbol{s}_R[v]) \\ &= \prod_{l=1}^{2U} \Phi(\sqrt{\rho}\beta_l(\boldsymbol{s}_R[v])). \end{aligned} \tag{73}$$

To phrase the detection problem as an optimization problem, the constraint on the eligibility set of the differential encoded information symbol is relaxed from the constellation set to the 2D real domain, i.e. $\boldsymbol{s}^{'}_R[v] \in \mathcal{S}_R$ is relaxed to $\boldsymbol{s}^{'}_R[v] \in \mathcal{R}^2$. With these, the likelihood can be optimized as follows

$$\begin{aligned} & \max_{\substack{\boldsymbol{s}^{'}_R[v]\in\mathcal{R}^2,\\ \left\|\boldsymbol{s}^{'}_R[v]\right\|^2=1.}} L(\boldsymbol{s}^{'}_R[v]) \\ = & \max_{\substack{\boldsymbol{s}^{'}_R[v]\in\mathcal{R}^2,\\ \left\|\boldsymbol{s}^{'}_R[v]\right\|^2=1.}} \prod_{i=1}^{2}\prod_{u=1}^{U} \Phi(\sqrt{\rho}\widetilde{\boldsymbol{f}}_{R,u,i}^T[v]\boldsymbol{s}^{'}_R[v]), \\ \overset{(a)}{\leq} & \max_{\substack{\boldsymbol{\beta}(\boldsymbol{s}^{'}_R[v])\in\mathcal{R}^{2U},\\ \left\|\boldsymbol{\beta}(\boldsymbol{s}^{'}_R[v])\right\|^2\leq\left\|\widetilde{\boldsymbol{F}}_R[v]\right\|^2.}} \prod_{l=1}^{2U} \Phi(\sqrt{\rho}\beta_l(\boldsymbol{s}^{'}_R[v])). \end{aligned} \tag{74}$$

where the inequality, (a), results from the Cauchy–Schwarz inequality. The optimization problem in the above equation is similar to the optimization problem derived in coherent literature for quantized distributed reception [6].

$$\begin{aligned} & \max_{\substack{\boldsymbol{\beta}(\boldsymbol{s}^{'}_R[v])\in\mathcal{R}^{2U},\\ \left\|\boldsymbol{\beta}(\boldsymbol{s}^{'}_R[v])\right\|^2\leq\left\|\widetilde{\boldsymbol{F}}_R[v]\right\|^2,\\ \boldsymbol{\beta}_l(\boldsymbol{s}^{'}_R[v])>0\forall\, l.}} \prod_{l=1}^{2U} \Phi(\sqrt{\rho}\beta_l(\boldsymbol{s}^{'}_R[v])) \\ = & \max_{\substack{\boldsymbol{\beta}(\boldsymbol{s}^{'}_R[v])\in\mathcal{R}^{2U},\\ \left\|\boldsymbol{\beta}(\boldsymbol{s}^{'}_R[v])\right\|^2=\left\|\widetilde{\boldsymbol{F}}_{R,t}\right\|^2,\\ \boldsymbol{\beta}_l(\boldsymbol{s}^{'}_R[v])>0\forall\, l.}} \prod_{l=1}^{2U} \Phi(\sqrt{\rho}\beta_l(\boldsymbol{s}^{'}_R[v])). \end{aligned} \tag{75}$$

The equality above follows trivially since the normal CDF is strict increasing as stated in Remark 2 below.

**Remark 2.** *While the feasible set in (75) is convex, the objective function is log-concave. The log-concavity of the objective function stems from the fact that the product of log-concave functions yields a log-concave function. Also, note that the objective function is strictly increasing over the domain $(0, \infty)$ and a log-concave function is also quasiconcave [32].*

**Remark 3.** $\beta_l(\boldsymbol{s}^{'}_R[v]) = \sqrt{\frac{1}{2U}}\left\|\widetilde{\boldsymbol{F}}_R[v]\right\|, \forall l$ *is an extreme point. This follows from the solution of the norm constraint in the feasibility set and the definition of an extreme point [33]. Also, since the objective function is strictly increasing in the interval $(0, \infty)$, the point $\beta_l(\boldsymbol{s}^{'}_R[v]) = \sqrt{\frac{1}{2U}}\left\|\widetilde{\boldsymbol{F}}_R[v]\right\|, \forall l$ is the only extreme point.*

**Lemma 1.** *The extreme point $\beta_l(\boldsymbol{s}^{'}_R[v]) = \sqrt{\frac{1}{2U}}\left\|\widetilde{\boldsymbol{F}}_R[v]\right\|, \forall l$ maximizes the likelihood in (75).*

*Proof.* To prove Lemma 1, we first notice that Remark (2) implies that the likelihood is also quasiconvex. This is due to the fact that every monotone function is both quasiconvex and quasiconcave. To find the maximum we use Theorem 3.5.3 in [34] - this states that the optimal solution to a quasiconvex problem exist at the extreme of its feasibility set. ☐

From Lemma 1, the vector $\boldsymbol{\beta}(\boldsymbol{s}^{'}_R[v]) = \sqrt{\frac{1}{2U}}\left\|\widetilde{\boldsymbol{F}}_R[v]\right\|\mathbf{1}_{2U}$ maximizes the likelihood function. Recall that, $\boldsymbol{\beta}(\boldsymbol{s}^{'}_R[v]) = \widetilde{\boldsymbol{F}}_R[v]\boldsymbol{s}^{'}_R[v]$, the information vector is

$$\boldsymbol{s}^{'}_R[v] = \widetilde{\boldsymbol{F}}^{\dagger}_R[v]\boldsymbol{\beta}(\boldsymbol{s}^{'}_R[v]) = \sqrt{\frac{1}{2U}}\left\|\widetilde{\boldsymbol{F}}_R[v]\right\|\widetilde{\boldsymbol{F}}^{\dagger}_R[v]\mathbf{1}_{2U}.$$

From the sign refinement operation in (65), the following relation exist between matrices $\widetilde{\boldsymbol{F}}_R[v]$ and $\boldsymbol{F}_R[v]$

$$\widetilde{\boldsymbol{F}}^{\dagger}_R[v]\mathbf{1}_{2U} = \boldsymbol{F}^{\dagger}_R[v]\boldsymbol{q}_R[v].$$

Hence, a reasonable detector for the information symbols is

$$\hat{\boldsymbol{x}}_R[v] = \boldsymbol{F}^{\dagger}_R[v]\boldsymbol{q}_R[v]. \tag{76}$$

This detector is similar to the coherent case [6] except that the $\boldsymbol{F}_R[v]$ matrix represents quantized symbols that were collected across all $U$ antennas at the previous time, while the matrix in the coherent case denotes the channel which is assumed to be known.

Finally, the symbol is detected as

$$\hat{a}^{'}[v]\hat{\boldsymbol{s}}_R[v] = \underset{(a^{'})^{'}[v]\boldsymbol{s}^{'}_R[v]\in\{\mathcal{A}\times S_R\}}{\arg\min}\left\|\hat{\boldsymbol{x}}_R[v] - (a^{'})^{'}[v]\boldsymbol{s}^{'}_R[v]\right\|^2. \tag{77}$$

For both ML and ID detectors, the bit $b_1[v]$ can be recovered using

$$\hat{b}_1[v] = \begin{cases} 0, & \text{if } \left\|\hat{a}^{'}[v]\hat{\boldsymbol{s}}_R[v]\right\| = 1, \\ \\ 1, & \text{if } \left\|\hat{a}^{'}[v]\hat{\boldsymbol{s}}_R[v]\right\| \neq 1, \end{cases} \tag{78}$$

and with an abuse of notation the remaining bits can be recovered from the phase information by $\{\hat{b}_2[v], \hat{b}_3[v], \cdots, \hat{b}_{N_b}[v]\} = \Upsilon^{-1}\left(\frac{\hat{a}^{'}[v]\hat{\boldsymbol{s}}_R[v]}{\left\|\hat{a}^{'}[v]\hat{\boldsymbol{s}}_R[v]\right\|}\right)$. Empirical results indicate that the detector suffers from a substantial error floor, which can be attributed to the performance of the amplitude recovery part of the detector. This is intuitive because the one-bit quantizer only represents one level of amplitude (i.e $1$ or $-1$).

### B. Higher Resolution Maximum Likelihood Receivers

In this section, the ML detector for quantized detection of DAPSk symbols is derived. If the quantizer is applied to the signal received at the $u-$th antenna, during the symbol interval $v$, the probability of obtaining a particular label, $q^R$, for the real part can be written as

$$\begin{aligned} & Pr(q^R|y_u[v] = a^{'}[v]y_u[v-1]s[v] + z^{'}_u[v]) \\ &= \Phi\left(\rho(\zeta_{l^R+1} - a^{'}[v]\sqrt{\rho}\boldsymbol{f}^T_{R,u,1}[v]\boldsymbol{s}_R[v])\right) \\ &- \Phi\left(\rho(\zeta_{l^R} - a^{'}[v]\sqrt{\rho}\boldsymbol{f}^T_{R,u,1}[v]\boldsymbol{s}_R[v])\right), \end{aligned} \tag{79}$$

where $l^R \in \{0, 1, \cdots, N_\zeta - 1\}$ and it satisfies $\zeta_{l^R} \leq y_{R,u,1}[v] < \zeta_{l^R+1}$. Similarly, $l^I \in \{0, 1, \cdots, N_\zeta - 1\}$ . A similar expression can be obtained for the imaginary part of the signal as

$$\begin{aligned} &Pr(q^I | y_u[v] = a^{'}[v] y_u[v-1] s[v] + z_u^{'}[v]) \\ &= \Phi\Bigg(\rho(\zeta_{l^I+1} - a^{'}[v]\sqrt{\rho}\boldsymbol{f}^T_{R,u,2}[v]\boldsymbol{s}_R[v])\Bigg) \\ &- \Phi\Bigg(\rho(\zeta_{l^I} - a^{'}[v]\sqrt{\rho}\boldsymbol{f}^T_{R,u,2}[v]\boldsymbol{s}_R[v])\Bigg), \end{aligned} \tag{80}$$

where $l^I \in \{0, 1, \cdots, N_\zeta - 1\}$ and it satisfies $\zeta_{l^I} \leq y_{R,u,2}[v] < \zeta_{l^I+1}$. Since, the real and imaginary samples are quantized independently, the probability of obtaining a complex label, $q = q^R + iq^I$ is

$$Pr(q | y_u[v]) = Pr(q^R | y_u[v]) Pr(q^I | y_u[v]), \tag{81}$$

and considering independent $U$ antennas, the probability function translates to

$$Pr(\boldsymbol{q} | \boldsymbol{y}[v]) = \prod_{u=1}^{U} Pr(\boldsymbol{q} | y_u[v]). \tag{82}$$

As discussed previously, there are three possible values of $\rho$ and a reasonable likelihood function can be written as

$$\begin{aligned} &L(a^{'}[v]\boldsymbol{s}_R[v] | \rho) \\ &= \max\{Pr(\boldsymbol{q}|\boldsymbol{y}[v]\rho_1), Pr(\boldsymbol{q}|\boldsymbol{y}[v]\rho_2), Pr(\boldsymbol{q}|\boldsymbol{y}[v]\rho_3)\}, \end{aligned} \tag{83}$$

following this line of reasoning, the maximum likelihood detector can be written as

$$\hat{a}^{'}[v]\hat{\boldsymbol{s}}_R[v] = \underset{(a^{'})^{'}[v]\boldsymbol{s}^{'}_R[v] \in \{\mathcal{A} \times S_R\}}{\arg\max} L((a^{'})^{'}[v]\boldsymbol{s}^{'}_R[v] | \rho). \tag{84}$$

The bit $b_1[v]$ can be recovered using (78) and with an abuse of notation the remaining bits can be recovered from the phase information by $\{\hat{b}_2[v], \hat{b}_3, \cdots, \hat{b}_{N_b}[v]\} = \Upsilon^{-1}\left(\frac{\hat{a}^{'}[v]\hat{\boldsymbol{s}}_R[v]}{\left\|\hat{a}^{'}[v]\hat{\boldsymbol{s}}_R[v]\right\|}\right)$.

### *C. Multi-Bit Receivers for Differential Amplitude Phase Shift Keying*

As stated in previous section, one-bit quantizers are unable to detect the amplitude change from symbol to symbol, hence, they are unsuitable for differential amplitude phase shift keying systems In this setup, we propose using two bits per real dimension for both amplitude and phase detection. The five decision thresholds for amplitude detection are specified as $\{\zeta_1 < \zeta_2 < \zeta_3 < \zeta_4 < \zeta_5\} = \{-\infty, \zeta_2, \zeta_3, \zeta_4, \infty\}$, the thresholds $\zeta_2, \zeta_3, \zeta_4$ can be thought of as optimization variables that can affect the decoding performance. Because information is conveyed by the changes in amplitude, a reasonable optimization constraint is to ensure that the quantization thresholds is dependent on the ring ratio, $a$. Except stated otherwise the following thresholds are used

$$\zeta_2 = -\cos\frac{\pi}{4}\sqrt{\frac{2a^2}{a^2+1}}, \zeta_3 = 0, \zeta_4 = \cos\frac{\pi}{4}\sqrt{\frac{2a^2}{a^2+1}}.$$

The amplitude can be obtained using (84), while the phase information can be recovered using the 1-bit maximum likelihood detector or inverse-decoding detector. Note that the exact quantization model used in (84) for amplitude detection is computational expensive to compute and only achieves reasonable BER at high SNR. Therefore, we develop differential detectors that avoid the use of the exact quantization models. The next section presents one of such receivers.

### *D. Differential Amplitude Detection with Low-Resolution ADC*

To develop a differential amplitude detector without the use of the exact quantization model, the Bussgang Theorem presented in (15) is employed to decompose the output of the non-linear quantizer into a desired component and an uncorrelated noise vector

$$\boldsymbol{q}[v] = \mathcal{Q}(\boldsymbol{y}[v]) = \eta \boldsymbol{I}_U \boldsymbol{y}[v] + \epsilon[v], \tag{85}$$

and employing (55), the quantized signal can be written as

$$\begin{aligned} \boldsymbol{q}[v] &= \eta \boldsymbol{h}[v] x[v] + \eta \boldsymbol{z}[v] + \epsilon[v], \\ \boldsymbol{q}[v] &= \eta \boldsymbol{h}[v] x[v] + \boldsymbol{\epsilon}^{'}[v], \end{aligned} \tag{86}$$

where $\boldsymbol{\epsilon}^{'}[v] = \eta\boldsymbol{z}[v] + \epsilon[v]$ is Gaussian, i.e $\boldsymbol{\epsilon}^{'}[v] \sim \mathcal{CN}(0, \tilde{\sigma}^2_\epsilon = \eta^2\sigma^2_z + \sigma^2_\epsilon)$ and considering a single antenna, $u$ in the real domain, we have

$$q_{R,u,i}[v] = \eta \boldsymbol{h}^T_{R,u,i}[v] \boldsymbol{x}_R[v] + \boldsymbol{\epsilon}^{'}_{R,u}[v], \tag{87}$$

where

$$\boldsymbol{\epsilon}^{'}_{R,u}[v] = \begin{bmatrix} (\epsilon^{'}_{R,u,1}[v]) \\ (\epsilon^{'}_{R,u,2}[v]) \end{bmatrix} = \begin{bmatrix} \Re(\epsilon^{'}_{R,u}[v]) \\ \Im(\epsilon^{'}_{R,u}[v]) \end{bmatrix}, \tag{88}$$

The second order statistics of the quantized received signal can be approximated as

$$\begin{aligned} \Lambda[v] = \frac{1}{U}\sum_{u=1}^{U} |q_{R,u,i}[v]|^2 = \tilde{x}^2[v]\eta^2 \frac{\sum_{u=1}^{U} \boldsymbol{h}^H_{R,u,i}[v]\boldsymbol{h}_{R,u,i}[v]}{U} \\ + \frac{\sum_{u=1}^{U} \epsilon^{'}_{R,u,i}[v]^H \epsilon^{'}_{R,u,i}[v]}{U}. \end{aligned} \tag{89}$$

**Remark 4.** *Asymptotically, the specific channel is not required because of channel hardening [35]. More specifically as $U \to \infty$, $\frac{\sum_{u=1}^{U} \boldsymbol{h}^H_{R,u,i}[v]\boldsymbol{h}_{R,u,i}[v]}{U}$ converges to a constant, $\alpha$. Also, the noise term converges to a constant, $\tilde{\sigma}^2_\epsilon = \eta^2\sigma^2_z + \sigma^2_\epsilon$.*

From remark 4, $\Lambda$ depends on the noise power, channel amplitude and the amplitude of the transmitted symbol, hence (87) can be written as

$$\Lambda[v] = \frac{1}{U}\sum_{u=1}^{U} |q_{R,u,i}[v]|^2 = \tilde{x}^2[v]\eta^2\alpha^2 + \tilde{\sigma}^2_\epsilon. \tag{90}$$

A maximum likelihood detection approach based on the observation $\Lambda$ is used to test the hypothesis that the symbol amplitude remains constant across adjacent symbols i.e, $\tilde{x}[v] = \tilde{x}[v-1]$. More specifically, a hypothesis testing rule can be used to determine if $b_1[v] = 1$ or if $b_1[v] = 0$. This hypothesis is defined as $\mathcal{H}_1$ and is confirmed if

$$\Omega(\Lambda | \mathcal{H}_1) > \Omega(\Lambda | \mathcal{H}_0), \tag{91}$$

where $\Omega(\Lambda|\mathcal{H}_1)$ is the conditional pdf of $\Lambda[v]$. Hence, the hypothesis test in (91) is used to develop an energy detection threshold between two neighbouring DAPSK concentric circles.

**Assumption 3.** *The channel amplitude, the quantization effect, and composite noise variance are known.*

Utilizing Assumption (3), the conditional pdf of $\Lambda$ follows a non-central chi-square distribution and can be written as

$$\Omega(\Lambda|\alpha,\tilde{x}[v],\eta,\tilde{\sigma}_{\epsilon}^2)=\frac{U}{\tilde{\sigma}_{\epsilon}^2}\left(\frac{\Lambda}{\alpha^2\tilde{x}^2[v]\eta^2}\right)^{U-1}e^{-\frac{U}{\tilde{\sigma}_{\epsilon}^2}(\Lambda+\alpha^2\tilde{x}^2[v]\eta^2)}$$
$$\boldsymbol{I}_{U-1}\left(\frac{2U}{\tilde{\sigma}_{\epsilon}^2}\sqrt{\Lambda\alpha^2\tilde{x}^2[v]\eta^2}\right), \tag{92}$$

where $U>0$ and $\boldsymbol{I}_{U-1}$ is the modified Bessel function of the first kind. While the distribution is dependent on the amplitude of the transmitted symbol, $\tilde{x}[v]\in\{\psi_0,\psi_1\}$, the respective distributions conditioned on either $\psi_0$ or $\psi_1$ are not symmetric. This asymmetry and presence of a first order Bessel function mandates that the detection threshold must be obtained numerically by determining the intersections points of the two resulting pdfs.

However, at large $U$, the central limit theorem ensures that $\Lambda$ is well approximated by a non-central Gaussian distribution, $\Lambda\sim\mathcal{CN}(\mu_\Lambda,\sigma_\Lambda^2)$ with mean and variance obtained from the mean and variance of (92)

$$\begin{aligned}\mu_\Lambda=&\tilde{\sigma}_{\epsilon}^2+\tilde{x}^2[v]\eta^2\alpha^2,\\ \sigma_\Lambda^2=&\frac{2\tilde{\sigma}_{\epsilon}^2}{U}\left(2\tilde{\sigma}_{\epsilon}^2+2\tilde{x}^2[v]\eta^2\alpha^2\right).\end{aligned} \tag{93}$$

The threshold is determined at the intersection of the two pdfs specified by $\tilde{x}[v]=\psi_0$ and $\tilde{x}[v]=\psi_1$ respectively. Hence, the detection threshold can be obtained by determing the positive square root of the quadratic problem

$$\begin{aligned}&[1/\sigma_{\Lambda,1}^2-1/\sigma_{\Lambda,0}^2]\gamma^2-2[\mu_{\Lambda,1}/\sigma_{\Lambda,1}^2-\mu_{\Lambda,0}/\sigma_{\Lambda,0}^2]\gamma+\\ &[\mu_{\Lambda,1}^2/\sigma_{\Lambda,1}^2-\mu_{\Lambda,0}^2/\sigma_{\Lambda,0}^2]+\log\frac{\sigma_{\Lambda,1}^2}{\sigma_{\Lambda,0}^2}=0,\end{aligned} \tag{94}$$

where $\mu_{\Lambda,l}$ and $\sigma_{\Lambda,l}^2$ is the mean and variance of the pdf generated by $\tilde{x}[v]=\psi_l$.

**Remark 5.** *From (90), a reasonable energy detector is*

$$\hat{\tilde{x}}[v]=\begin{cases}\psi_0, & \textit{if } \Lambda[v]<\gamma,\\ \psi_1, & \textit{Otherwise.}\end{cases} \tag{95}$$

To determine bit $\hat{b}_1[v]$, we have to detect a change in the amplitude from one symbol to the next. More specifically,

$$\hat{b}_1[v]=\begin{cases}0, & \text{if } \hat{\tilde{x}}[v]=\hat{\tilde{x}}[v-1],\\ 1, & \text{Otherwise.}\end{cases} \tag{96}$$

The phase information is detected using either the maximum likelihood or inverse-decoding detectors, subsequently the remaining bits can be obtained as $\{\hat{b}_2[v],\hat{b}_3[v],\cdots,\hat{b}_{N_b}[v]\}=\Upsilon^{-1}(\hat{\boldsymbol{s}}_R[v]^)$.

### E. One Bit ADCs with Variable Quantization Levels

In the previous section considering multi-bit detection, 2 bits are used to detect both the amplitude and the phase information. Although, this falls under the category of low-resolution ADC, it is plausible to reduce the number of quantization bits by grouping antennas such that each antenna group employs quantizers with different levels. In this work, we propose three quantization based antenna groups specified as $\mathcal{U}_1$,$\mathcal{U}_2$, $\mathcal{U}_3$, such that $|\mathcal{U}_1|+|\mathcal{U}_2|+|\mathcal{U}_3|=U$. The quantization function operates on each group as follows

$$q_{R,u_j,i}[v]=\mathcal{Q}(y_{R,u_j,i}[v]), \tag{97}$$

where $u_j\in\mathcal{U}_j$, and $y_{R,u_j,i}[v]\in\{\zeta_{1,j},\zeta_{3,j}\}$. More specifically,

$$q_{R,u_j,i}[v]=\begin{cases}\zeta_{3,j}, & \text{if } y_{R,u_j,i}[v]>\zeta_{2,j},\\ \\ \zeta_{1,j}, & \text{if } y_{R,u_j,i}[v]<\zeta_{2,j}.\end{cases}$$

Utilizing the definitions in the previous sections (see 87), the quantization operation can be converted into a linear operation as follows

$$q_{R,u_j,i}[v]=\eta_j\boldsymbol{h}_{R,u_j,i}^T[v]\boldsymbol{x}_R[v]+\boldsymbol{\epsilon}_R^{'}[v], \tag{98}$$

The second order statistics of the quantized received signal can be approximated as

$$\begin{aligned}&\Lambda=\frac{1}{U}\sum_{u=1}^{U}|q_{R,u_j,i}[v]|^2=\\ &\tilde{x}^2[v]\frac{\sum_{j=1}^3\sum_{u\in\mathcal{U}_j}\eta_j^2\boldsymbol{h}_{R,u,i}^H[v]\boldsymbol{h}_{R,u,i}[v]}{U}\\ &+\frac{\sum_{j=1}^3\sum_{u\in\mathcal{U}_j}\epsilon_{R,u,i}^{'}[v]^H\epsilon_{R,u,i}^{'}[v]}{U}.\end{aligned} \tag{99}$$

**Assumption 4.** *We assume that quantization gains and quantization noise variance are equal across all antenna groupings.*

From Assumption 4 and Remark 4, the quantized received signal can be written similar to (90)

$$\Lambda[v]=\frac{1}{U}\sum_{u=1}^{U}|q_{R,u,i}[v]|^2=\tilde{x}^2[v]\eta^2\alpha^2+\tilde{\sigma}_{\epsilon}^2. \tag{100}$$

Hence, the amplitude ratio detector proposed in Remark 5 can be used, and the left most bit can be detected using (96). However, to employ the 1-bit detectors proposed in (69) and (77) for phase detection, we set $a^{'}[v]=1$, and assign a particular group to use the sgn function as its quantizer. If the $j-$th group employs a 1-bit signum quantizer, then $\boldsymbol{q}_{R,u_j,i}[v]\in\{-1,1\}$, and $m_{2,j}=0$. Hence, the maximum likelihood detector can be written as

$$\begin{aligned}\hat{\boldsymbol{s}}_R[v]&=\underset{\boldsymbol{s}_R^{'}[v]\in S_R}{\arg\max}\,L(\boldsymbol{s}_R^{'}[v]|\rho),\\ &=\underset{\boldsymbol{s}_R^{'}[v]\in S_R}{\arg\max}\prod_{i=1}^{2}\prod_{u\in\mathcal{U}_j}\Phi\left(\sqrt{\rho}\widetilde{\boldsymbol{f}}_{R,u,i}^T[v]\boldsymbol{s}_R^{'}[v]\right).\end{aligned} \tag{101}$$

Similarly, the inverse-decoding detector can be obtained for phase detection by utilizing the received signals from the $j-$th group.

### *F. Receiver Performance*

The previous section shows that the amplitude of the estimated symbol converges to the true amplitude corrupted by a scalar and an additive component. In this section, we analyze the asymptotic performance of the maximum likelihood phase detector presented in (67).

**Lemma 2.** *If $\check{s}_{R,ML}$ is the output of the ML detector presented in (67) with $a^{'}[v]=1$*

$$\check{\boldsymbol{s}}_{R,ML} = \underset{\substack{\boldsymbol{s}^{'}_R\in\mathbb{R}^2,\\ \left\|\boldsymbol{s}^{'}_R\right\|^2=1.}}{\arg\max}\, L(\boldsymbol{s}^{'}_R),$$

*then in probability, for high SNR $\rho \gg 1$, $\check{\boldsymbol{s}}_{R,ML}$ converges to the true differentially encoded symbol, $\boldsymbol{s}_R$. More specifically, as $U \to \infty$ then $\check{\boldsymbol{s}}_{R,ML} \xrightarrow{p} \boldsymbol{s}_R$.*

*Proof.* To proof this lemma, we consider any vector $\boldsymbol{m}_R \in \mathbb{R}^2 \backslash \boldsymbol{s}_R$, we need to show that in probability

$$L(\boldsymbol{s}_R) > L(\boldsymbol{m}_R),$$

as $U \to \infty$ with the constraint $\|\boldsymbol{m}_R\|^2 = 1$. The logarithm function is expanded as

$$L(\boldsymbol{s}^{\dagger}_R) = \sum_{i=1}^{2}\sum_{u=1}^{U} \log\Phi\left(\sqrt{\rho}\,\widetilde{\boldsymbol{f}}^{T}_{R,u,i}[v]\boldsymbol{s}^{\dagger}_R\right).$$

Since, the quantized signal received at the previous symbol duration, $\widetilde{\boldsymbol{f}}_{R,u,i}$ is independent across all $U$ and across all $i$, we have

$$\lim_{U\to\infty}\frac{1}{U}\sum_{u=1}^{U}\log\Phi\left(\sqrt{\rho}\,\widetilde{\boldsymbol{f}}^{T}_{R,u,i}[v]\boldsymbol{s}^{\dagger}_R\right) \xrightarrow{p} \mathbb{E}\left[\log\Phi\left(\sqrt{\rho}\,\widetilde{\boldsymbol{f}}^{T}_{R,u,i}[v]\boldsymbol{s}^{\dagger}_R\right)\right],$$

the weak law of large numbers allow for

$$\frac{1}{U}L(\boldsymbol{s}^{\dagger}_R) \xrightarrow{p} 2\mathbb{E}\left[\log\Phi\left(\sqrt{\rho}\,\widetilde{\boldsymbol{f}}^{T}_{R,u,i}[v]\boldsymbol{s}^{\dagger}_R\right)\right].$$

Finally, we have to show that

$$\mathbb{E}\left[\log\Phi\left(\sqrt{\rho}\,\widetilde{\boldsymbol{f}}^{T}_{R,u,i}[v]\boldsymbol{s}^{\dagger}_R\right)\right] > \mathbb{E}\left[\log\Phi\left(\sqrt{\rho}\,\widetilde{\boldsymbol{f}}^{T}_{R,u,i}[v]\boldsymbol{m}_R\right)\right],$$

Due to remark 2, this is equivalent to showing $\widetilde{\boldsymbol{f}}^{T}_{R,u,i}[v]\boldsymbol{s}^{\dagger}_R \overset{d}{>} \widetilde{\boldsymbol{f}}^{T}_{R,u,i}[v]\boldsymbol{m}_R$, where $\overset{d}{>}$ represents the first-order stochastic dominance. Since, all $\widetilde{\boldsymbol{f}}^{T}_{R,u,i}[v]$ are independent, this stochastic dominance can be proved by following Appendix B in [6]. □

## V. Numerical Results

In this section, we evaluate the performance of the proposed low-resolution differential detectors with Monte-Carlo simulations. We consider both single-carrier and multi-carrier systems. We employ the exponential channel model to generate the PDP. We assume a sampling rate of $T_s = 50 \times 10^{-9}$ and the three different values of the relative delay spreads, $\tau_{trms} = \{50, 100, 150\} \times 10^{-9}$. These delay spreads produce channels of different lengths $L = \{11, 21, 31\}$ and these values correspond to different degrees of frequency selectivity. In addition to these PDP values, the multi-carrier channel is assumed to have a Doppler spread and to take on one of the following values $f_l = \{5, 50, 5000\}$ Hz. To provide a baseline, we compare the proposed differential detectors to the maximum likelihood one-bit detectors provided in [6], [14]. Those works assume that the channel remains stationary during the entire blocks of transmission. In frequency or fast fading scenarios, these assumptions are not valid. In this paper, the pilot sequences used for comparison are assumed to occupy a fraction of the total number of channel uses in the single carrier system or a fraction of the number of subcarriers in the multi-carrier system. The following fraction of channel uses are considered $\xi = \{12.5\%, 25\%, 50\%\}$. The quantized received sequences are extracted and a least-squares approach is used to generate channel estimates for the data symbols. In the differential system, the first $N_s$ transmitted symbols can be viewed as reference symbols. This is because they are redundant as they carry no information. Without loss of generality, the results presented consider two transmit antennas $K = 2$, a differential encoding block size $N_s = 2$, $N = 256$ channel uses for the single carrier system, and $N = 256$ subcarriers for the OFDM system. Hence, the fraction of reference signals used in the differential system occupies less than $0.8\%$ of the available channel uses or available OFDM subcarriers. The spectral efficiency of the proposed detectors is calculated as

$$S.E. = \begin{cases} \xi N N_b(1 - SER), & \text{if } SER \geq SER_{th}, \\ 0, & \text{otherwise}, \end{cases}$$

where $SER_{th}$ is a threshold of the symbol error rate which is derived from the block error rate. In this work, this $SER_{th}$ value is set to $5\%$.

### *A. Complexity analysis.*

The complexity of the ML detector presented in (40) is exponential in both $M$ and $N_s$. This detector requires an exhaustive search over $M^{N_s}$ possible symbols. Also, the normal CDF in this detector requires a high degree of precision, this necessities the use of highly precise numerical algorithms. The equivalent detector presented in (53) consist of $N_s$ parallel detectors each with a linear complexity in $M$. The ML detector presented in (69) for DAPSK symbol detection has a complexity of $(|\mathcal{X}|M)^{N_s}$. Although the DAPSK system presented focuses on $N_s = 1$, it can easily be extended for $N_s > 1$. The alternate ID detector circumvents the need for the normal CDF and is linear $M$.

While the detectors presented in this paper are similar in complexity with the detectors presented in the coherent literature [6], [14], coherent detectors generate additional complexity due to channel estimation and equalization. More specifically, the least squares technique used to generate channel estimate for the data symbols require additional matrix multiplication and matrix inversion. Also, for frequency selective channels the number of paths $L$ needs to be estimated [36].

### *B. Evaluation of the proposed low-resolution detector for DPSK systems*

We present the performance of the detector presented in (53) considering frequency selective channels. Figure (1a) indicate an improvement in the BER performance as the number of receive antennas increases for both the differential and coherent schemes when 8-DPSK is employed. The differential scheme has a similar BER performance to the coherent scheme that uses $\xi = 0.125$ fraction of the channel uses for pilot transmission. The coherent scheme with $\xi = 0.5$ outperforms both schemes in terms of the BER albeit at the cost of spectral efficiency. The benefit of the differential scheme is noticeable in the spectral efficiency plot shown in Figure (1b). The differential scheme outperforms both coherent schemes when the SNR is above zero.

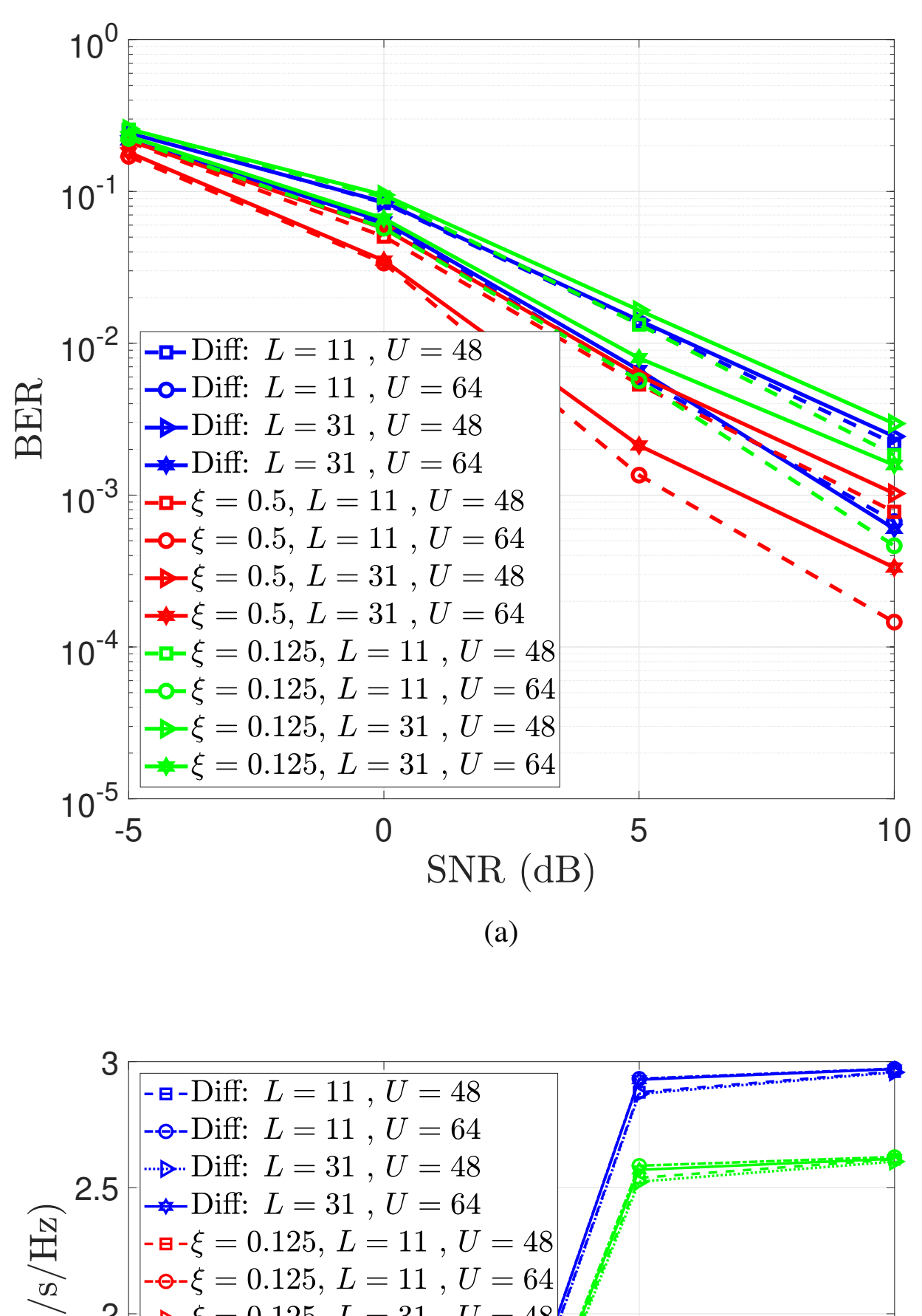

(a)

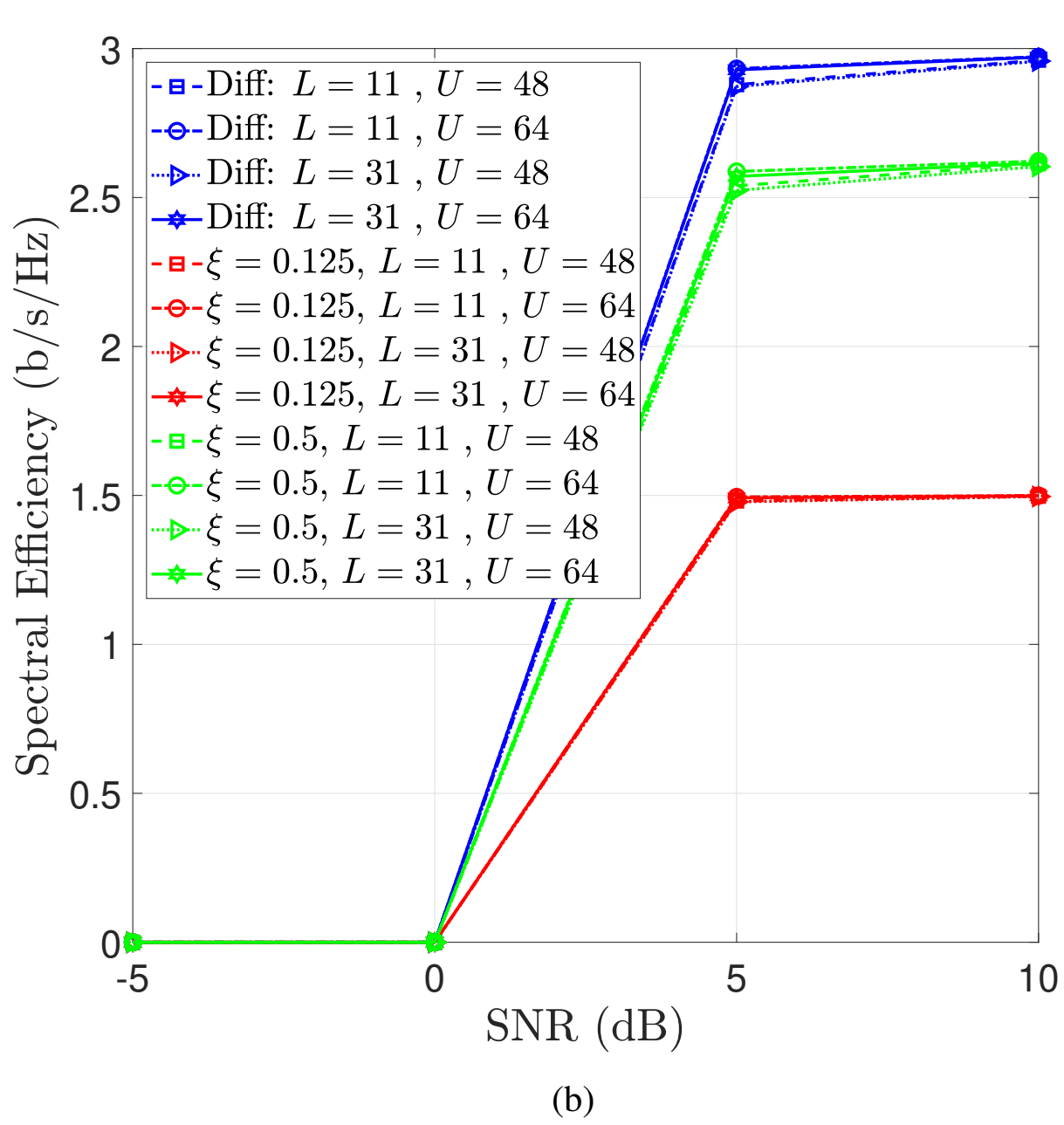

(b)

Figure 1. (a) BER and (b) spectral efficiency for 8-DPSK in the differential system and 8-PSK in the coherent scheme.

For 16-DPSK, the BER also improves as the number of receive antennas increases. Similar to 8-DPSK, the BER performance of the differential scheme and the BER performance of the coherent schemes with $\xi = 0.125$ are comparable across all channel conditions. The BER of the coherent scheme with $\xi = 0.5$ is lower than the BER performance of the differential scheme across all channel conditions. From Figure (2b), for both the coherent scheme with $\xi = 0.125$ and the differential scheme, the spectral efficiency is very dependent on the number of receive antennas. For $U = 128$ antennas, the spectral efficiency of both schemes is only non-zero above 5 dB while for $U = 256$ antennas the spectral efficiency is non-zero above 0 dB.

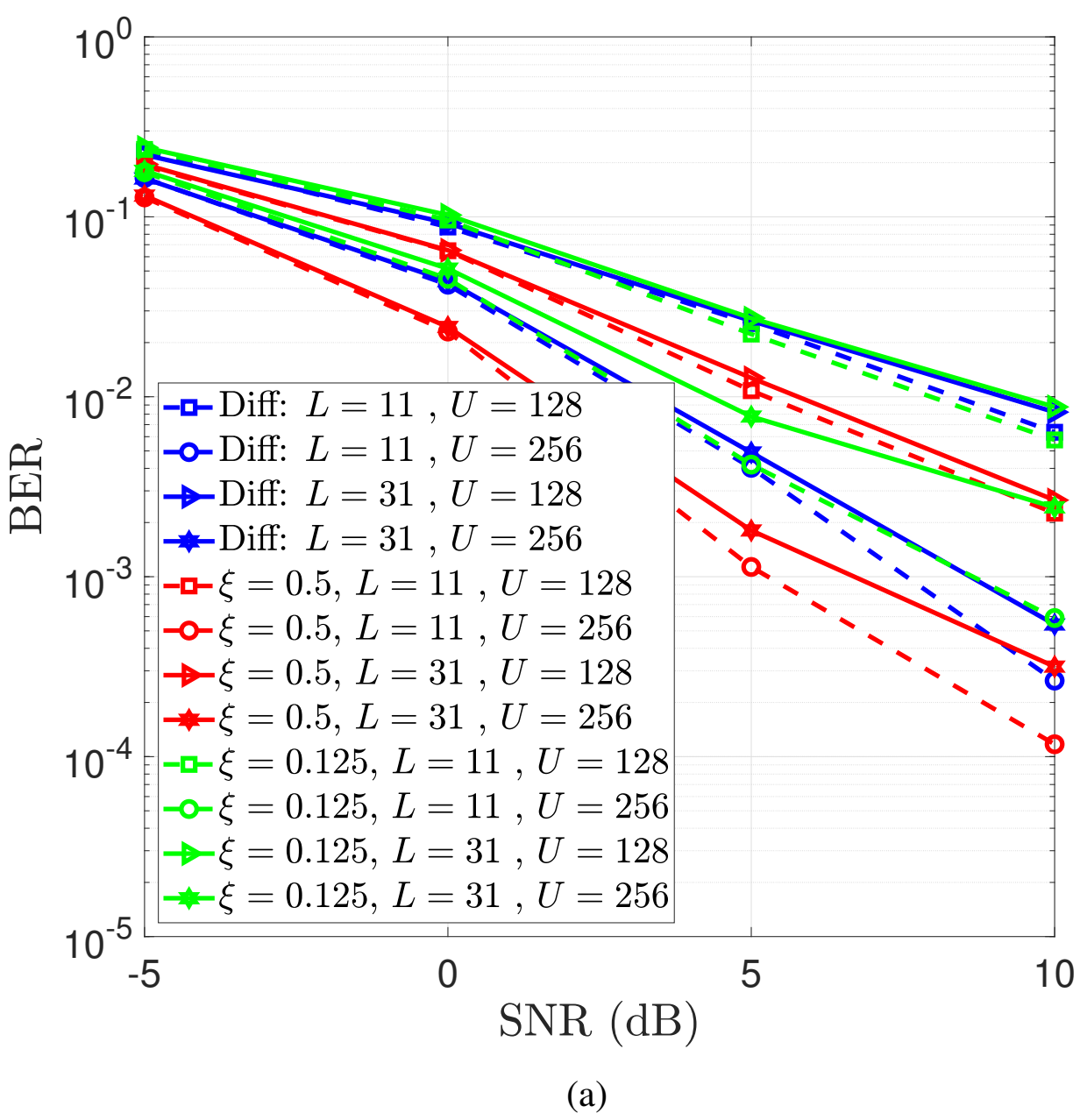

(a)

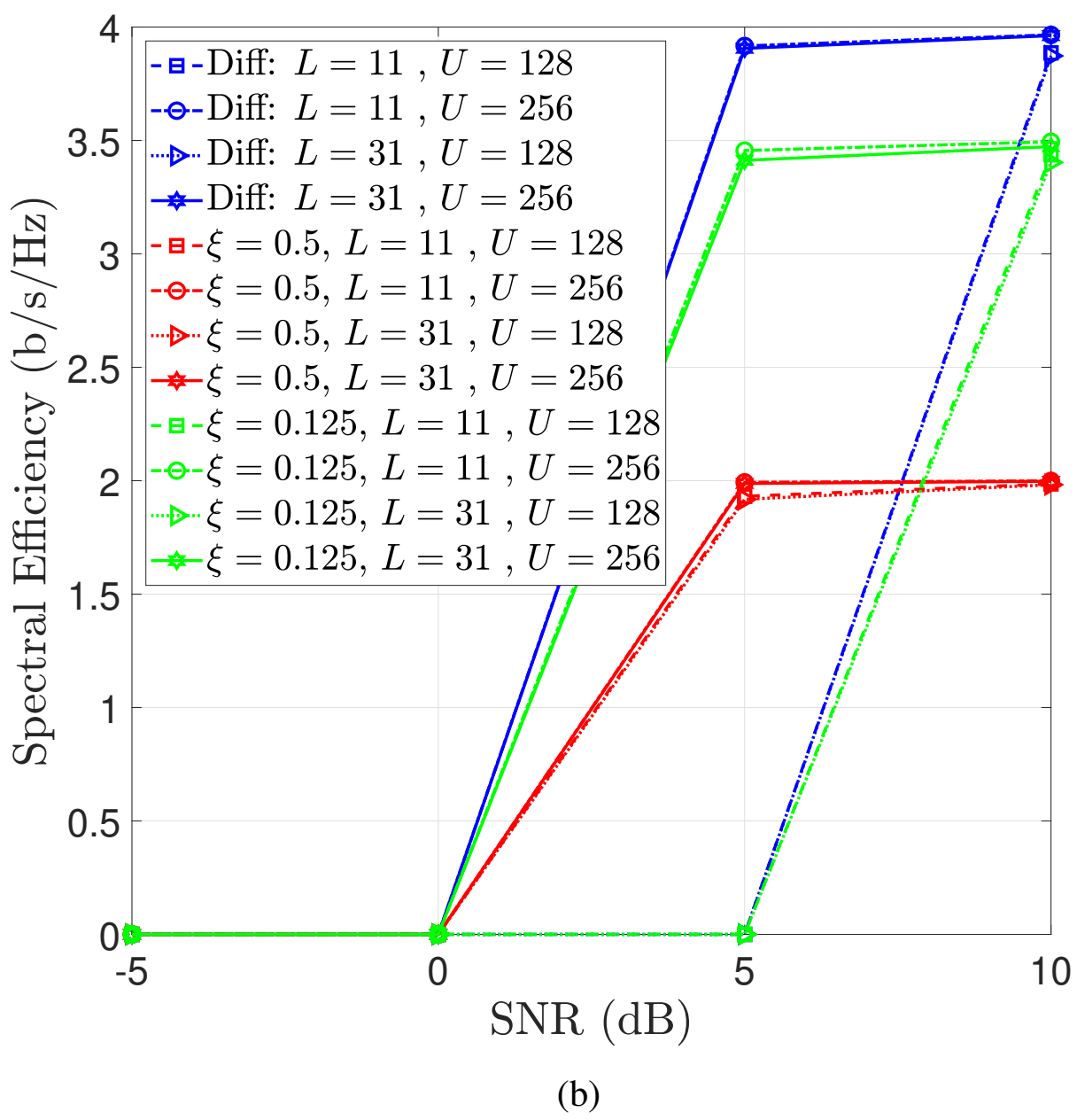

(b)

Figure 2. (a) BER and (b) spectral efficiency for 16-DPSK in the differential system and 16-PSK in the coherent scheme.

The BER for the OFDM system is shown in Figure (3) under slow and fast fading conditions. The channels are also frequency selective and the differential scheme outperforms the low-resolution coherent scheme with $\xi = 0.125$. For $U = 64$ antennas with $L = 21$ channel taps and $f_d = 50$, the differential scheme shows a 1 dB improvement over the coherent schemes. For the fast fading condition $f_d = 5000$, this margin increases to 2.5 dB. An increase in the number of antennas also results in an increase in these margins.

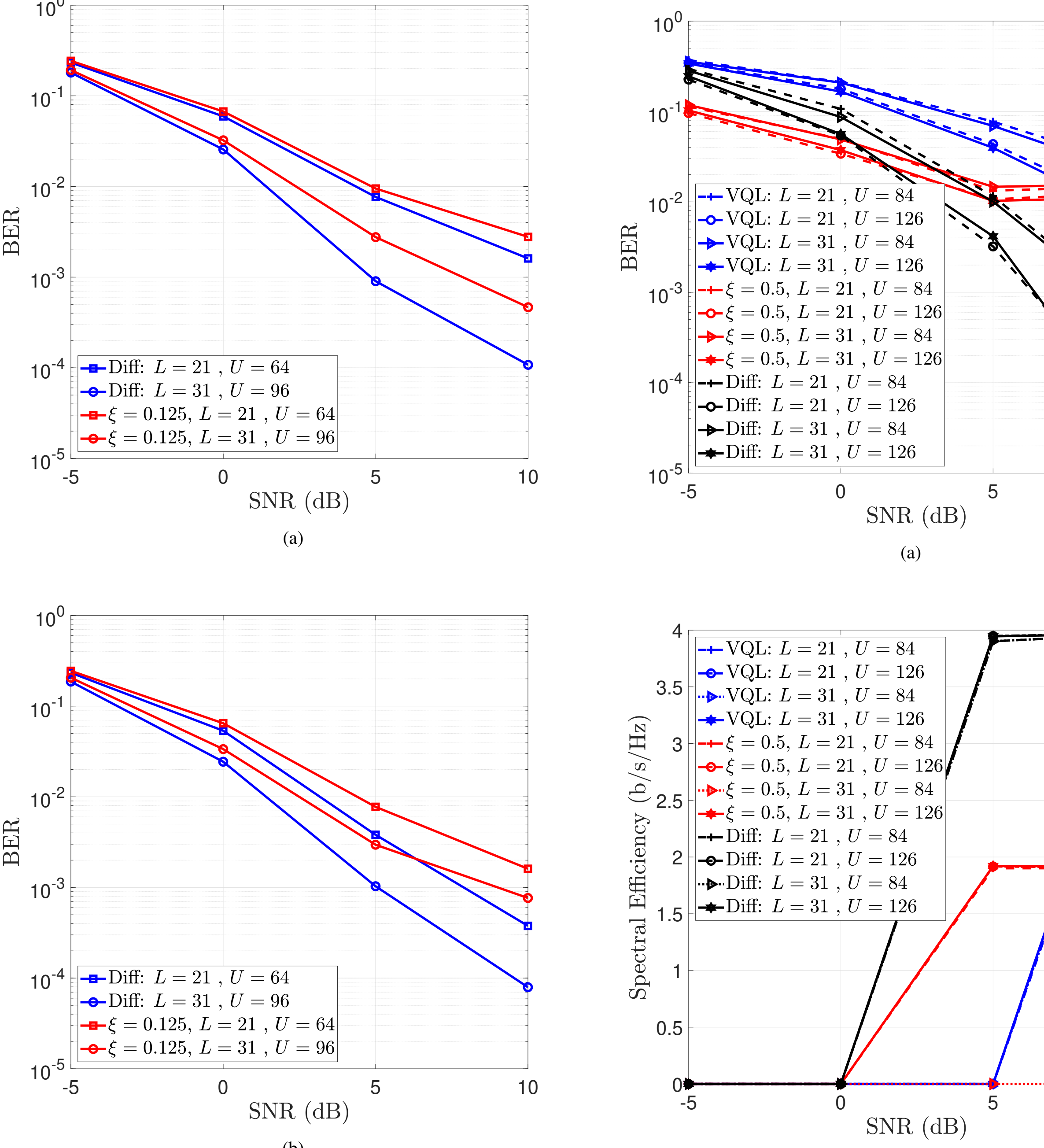

Figure 3. BER curves for 4-DPSK in the differential OFDM scheme and QPSK in the coherent OFDM scheme with (a) $f_d = 50$ Hz and (b) $f_d = 5$ KHz.

Figure 4. (a) BER and (b) spectral efficiency for 16-DAPSK in the differential scheme using the ML detector and 16-QAM in the coherent scheme.

### *C. Evaluation of the proposed low-resolution detector for DAPSK*

In this section, we focus on the detection of DAPSK systems in single-carrier systems. Although, the threshold design for the energy detector is specified in (93), the thresholds used in this section are obtained through Monte-Carlo simulations. These thresholds depend on the number of antennas and the SNR. The low-resolution and the variable quantization level detectors both employ (95) to detect the amplitude information. For both detectors, the phase information can be detected using either the ML detectors (69) or the ID detectors (77).

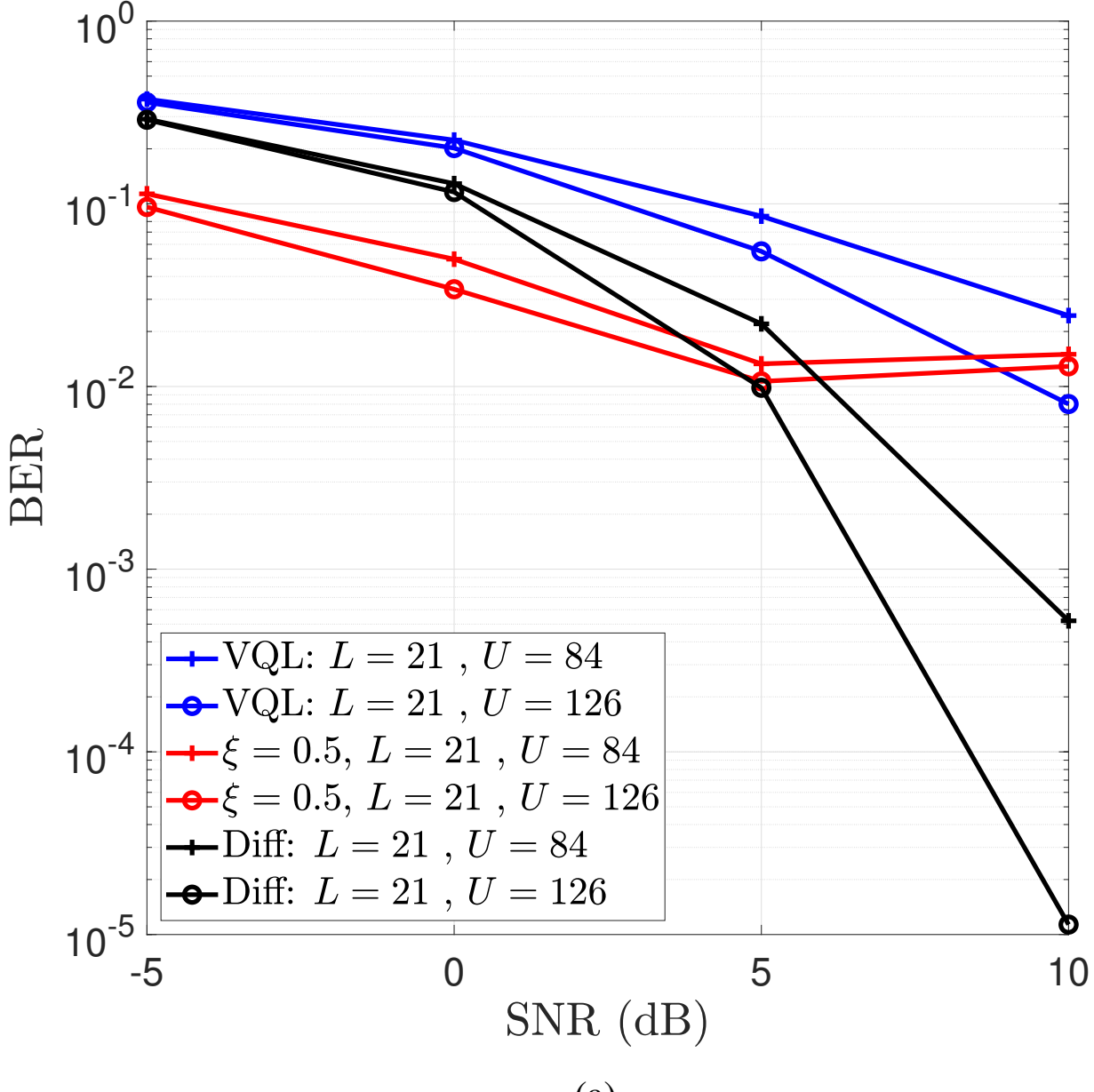

(a)

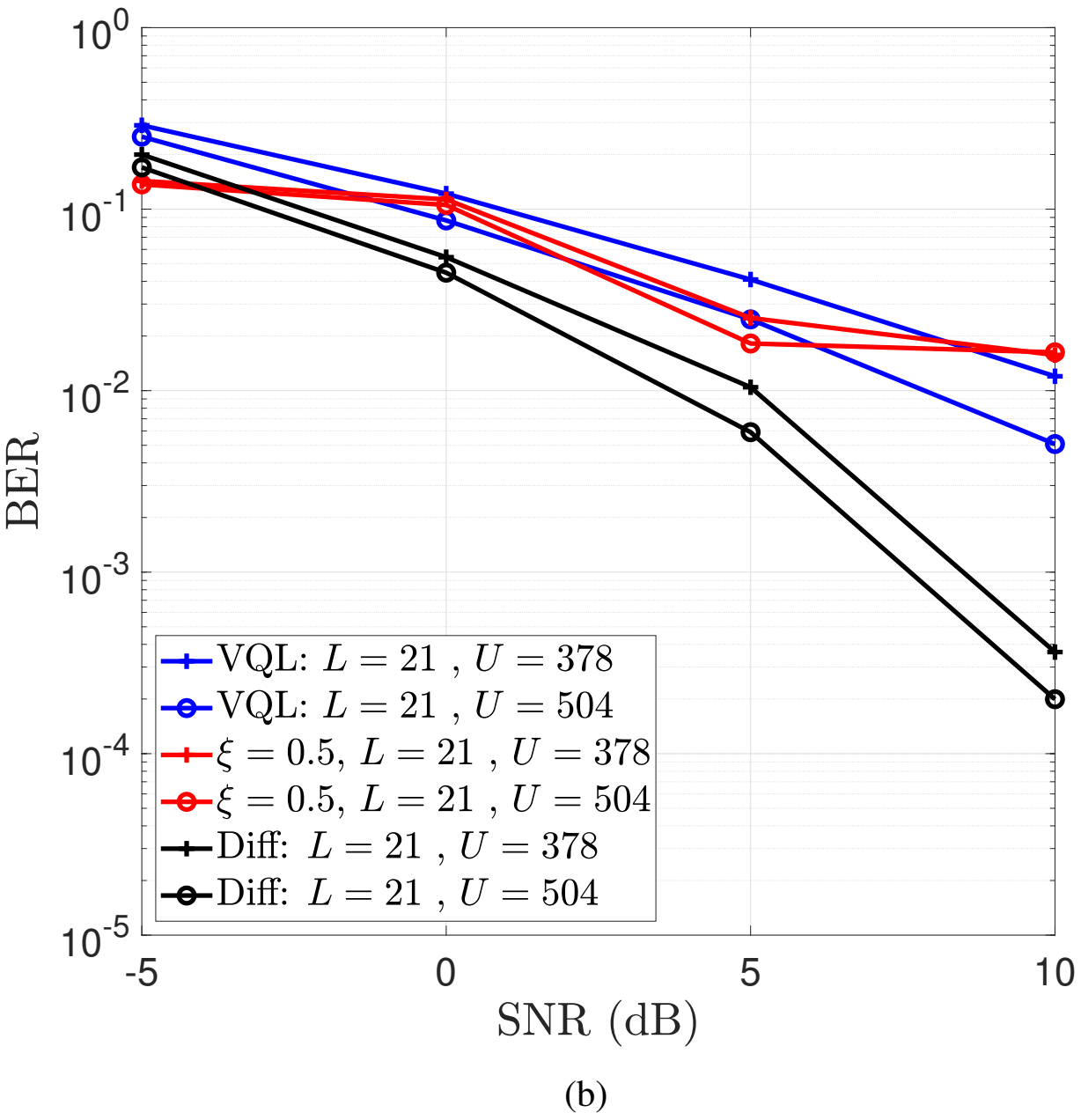

(b)

Figure 5. (a) BER for 16-DAPSK with the ID decoder (b) BER for 32-DAPSK with the ML decoder.

Note that the amplitude detection employing the low-resolution energy detector is termed "Diff" while the amplitude detection employing variable quantization levels is termed "VQL"[7].

[7]Note that the amplitude detection employing the low-resolution energy detector is termed "Diff" while the amplitude detection employing variable quantization levels is termed "VQL".

Figure (4) provides the maximum likelihood detection performance of 16-DAPSK in terms of BER and spectral efficiency.

The BER performance of both differential schemes is improved with an increase in the number of antennas. At low SNR, the coherent scheme outperforms both the VQL and the Diff detectors. For $U = 84$, the Diff detector outperforms the coherent detectors above an SNR of $4$ dB. While for $U = 126$, the Diff detector outperforms the coherent detectors above an SNR of $1$ dB. Above $0$ dB, the Diff detector always outperforms both the coherent scheme and the VQL schemes in terms of spectral efficiency.

Figure (5a) and (5b) present BER performance curves for the inverse decoding of 16 DAPSK symbols and the maximum likelihood decoding of 32-DAPSK symbols. The coherent detector becomes limited in the high SNR regime, this is in-line with the coherent detectors proposed in [6].

## VI. Conclusion

This paper has investigated the use of differential modulation at a transmitter and a large number of antennas at the BS with each antenna having low-resolution ADCs. First, we considered a differential phase shift keying system with square differential orthogonal matrices. The Bussganag theorem is used to express the quantized received signal in terms of quantized signals received during previous channel uses. Utilizing these expressions, low-resoultion detectors are developed and evaluated in both single carrier systems and OFDM systems. For differential amplitude phase shift keying systems, we developed low-resolution energy detector for detecting the amplitude information. We show that the threshold is asymptotically dependent on the SNR. We also develop maximum likelihood and inverse decoding detectors for detecting the phase information. Finally, we show through Monte-Carlo simulations that the designed detectors achieve reasonable BER in the large antenna regime and they are resilient to changes in the channel. These differential detectors also provide better throughput than coherent schemes since the coherent detector require substantial overhead for channel estimation.

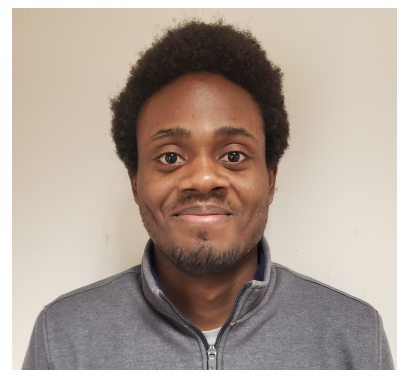

**Don-Roberts Emenonye** (Graduate Student Member, IEEE) received the B.Sc. degree in electrical and electronics engineering from the University of Lagos, Nigeria, in 2016, and the M.S. degree in electrical engineering from Virginia Tech in 2020. He is currently pursuing the Ph.D. degree with the Bradley Department of Electrical and Computer Engineering, Virginia Tech, USA. His research interests span the area of communication theory and wireless positioning.

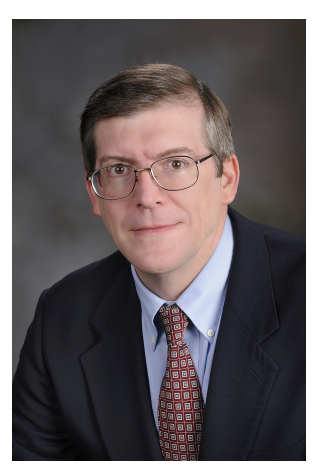

**Carl Dietrich** (SM’13) earned Electrical Engineering degrees from Virginia Tech (Ph.D. & M.S.) and Texas A&M University (B.S.). His research interests are in multiple aspects of wireless communication systems including cognitive radio testing, radio wave propagation, and multi-antenna systems. He is a member of IEEE HKN and ASEE and a professional engineer in Virginia.

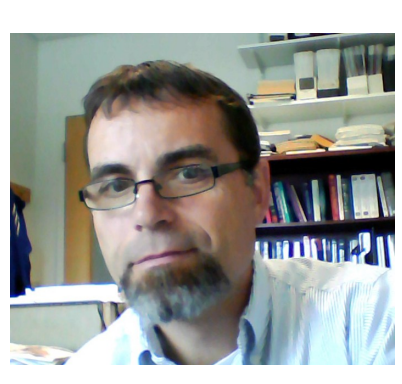

**Dr. R. Michael Buehrer** (IEEE Fellow 2016) joined Virginia Tech from Bell Labs as an Assistant Professor with the Bradley Department of Electrical and Computer Engineering in 2001. He is currently a Professor of Electrical Engineering and is the director of Wireless @ Virginia Tech, a comprehensive research group focusing on wireless communications. During 2009 Dr. Buehrer was a visiting researcher at the Laboratory for Telecommunication Sciences (LTS) a federal research lab which focuses on telecommunication challenges for national defense. While at LTS, his research focus was in the area of cognitive radio with a particular emphasis on statistical learning techniques.

Dr. Buehrer was named an IEEE Fellow in 2016 "for contributions to wideband signal processing in communications and geolocation." His current research interests include machine learning for wireless communications and radar, geolocation, position location networks, cognitive radio, cognitive radar, electronic warfare, dynamic spectrum sharing, communication theory, Multiple Input Multiple Output (MIMO) communications, spread spectrum, interference avoidance, and propagation modeling. His work has been funded by the National Science Foundation, the Defense Advanced Research Projects Agency, Office of Naval Research, the Army Research Lab, the Air Force Research Lab and several industrial sponsors.

Dr. Buehrer has authored or co-authored over 80 journal and approximately 250 conference papers and holds 18 patents in the area of wireless communications. In 2010 he was co-recipient of the Fred W. Ellersick MILCOM Award for the best paper in the unclassified technical program. He is currently an Area Editor IEEE Wireless Communications. He was formerly an associate editor for IEEE Transactions on Communications, IEEE Transactions on Vehicular Technologies, IEEE Transactions on Wireless Communications, IEEE Transactions on Signal Processing, IEEE Wireless Communications Letters, and IEEE Transactions on Education. He has also served as a guest editor for special issues of The Proceedings of the IEEE, and IEEE Transactions on Special Topics in Signal Processing. In 2003 he was named Outstanding New Assistant Professor by the Virginia Tech College of Engineering and in 2014 he received the Dean's Award for Excellence in Teaching.